\documentclass[preprint]{aastex}

\usepackage{graphicx}
\usepackage{epstopdf}
\usepackage{lineno}

\newcommand{\Fermi}{\textit{Fermi}}

\shorttitle{\Fermi\ GBM Gamma-Ray Burst Catalog}
\shortauthors{Paciesas et al.}

\begin{document}

\title{The \Fermi\ GBM Gamma-Ray Burst Catalog: \\
  The First Two Years}

\author{
William S. Paciesas\altaffilmark{1},
Charles A. Meegan\altaffilmark{2},
Andreas von Kienlin\altaffilmark{3},
P. N. Bhat\altaffilmark{1},
Elisabetta Bissaldi\altaffilmark{4},
Michael S. Briggs\altaffilmark{1},
J. Michael Burgess\altaffilmark{1}, 
Vandiver Chaplin\altaffilmark{1}, 
Valerie Connaughton\altaffilmark{1}, 
Roland Diehl\altaffilmark{3}, 
Gerald J. Fishman\altaffilmark{5},
Gerard Fitzpatrick\altaffilmark{6}, 
Suzanne Foley\altaffilmark{3}, 
Melissa Gibby\altaffilmark{7}, 
Misty Giles\altaffilmark{7}, 
Adam Goldstein\altaffilmark{1},
Jochen Greiner\altaffilmark{3}, 
David Gruber\altaffilmark{3}, 
Sylvain Guiriec\altaffilmark{1}, 
Alexander J. van der Horst\altaffilmark{2}, 
R. Marc Kippen\altaffilmark{8}, 
Chryssa Kouveliotou\altaffilmark{5}, 
Giselher Lichti\altaffilmark{3},
Lin Lin\altaffilmark{1},
Sheila McBreen\altaffilmark{6}, 
Robert D. Preece\altaffilmark{1},
Arne Rau\altaffilmark{3}, 
Dave Tierney\altaffilmark{6} 
and 
Colleen Wilson-Hodge\altaffilmark{5}
}
  
\altaffiltext{1}{Center for Space Plasma and Aeronomic Research, University of Alabama in Huntsville, 320 Sparkman Drive, Huntsville, AL 35805, USA}
\altaffiltext{2}{Universities Space Research Association, 320 Sparkman Drive, Huntsville, AL 35805, USA}
\altaffiltext{3}{Max-Planck-Institut f\"{u}r extraterrestrische Physik, Giessenbachstrasse 1,
85748 Garching, Germany}
\altaffiltext{4}{Institute of Astro and Particle Physics, University Innsbruck, Technikerstrasse 25,
6176 Innsbruck, Austria}
\altaffiltext{5}{Space Science Office, VP62, NASA/Marshall Space Flight Center, Huntsville,
AL 35812, USA}
\altaffiltext{6}{School of Physics, University College Dublin, Belfield, Stillorgan Road, Dublin 4, Ireland}
\altaffiltext{7}{Jacobs Technology, Inc., 1525 Perimeter Pkwy NW, Huntsville, AL 35806, USA}
\altaffiltext{8}{Los Alamos National Laboratory, PO Box 1663, Los Alamos, NM 87545, USA}

\begin{abstract}
The \Fermi\ Gamma-ray Burst Monitor (GBM) is designed to enhance the scientific return from \Fermi\ in studying gamma-ray bursts (GRBs). In its first two years of operation GBM triggered on 491 GRBs. We summarize the criteria used for triggering and quantify the general characteristics of the triggered GRBs, including their locations, durations, peak flux, and fluence. This catalog is an official product of the \Fermi\ GBM science team, and the data files containing the complete results are available from the High-Energy Astrophysics Science Archive Research Center (HEASARC).

\end{abstract}
 

\section{Introduction}

The \Fermi\ Gamma-ray Space Telescope was launched on 11~June 2008 on a mission to study the universe at high energies. One of \Fermi's highest priority objectives is to help explain the physical mechanisms responsible for the powerful high-energy emission from gamma-ray bursts (GRBs). The \Fermi\ Gamma-ray Burst Monitor (GBM) supports that goal by detecting and measuring the prompt emission from GRBs and providing quick notification to \Fermi's main instrument, the Large Area Telescope (LAT), as well as to ground-based observers. The broad energy coverage of the GBM (8~keV -- 40~MeV) and the LAT (30~MeV -- 300~GeV) combine to measure the emission spectra of sufficiently bright GRBs over more than seven decades of energy.

The on-board GBM trigger system for detecting GRBs was first enabled on 12 July 2008. In this paper we provide a catalog of GRBs that triggered the GBM during its first two years of operation. During this time the instrument burst detection algorithms were triggered 908 times: 492 of these are classified as GRBs, 79 as terrestrial gamma-ray flashes (TGFs), 170 as soft gamma-ray repeaters (SGRs), 31 as solar flares, 61 as charged particles and 75 as others (galactic sources, accidental statistical fluctuations, or too weak to classify). Of the 491 GRBs (in one case the same GRB triggered GBM twice), 18 were detected by the LAT with high confidence above 100~MeV  \citep{LAT_cat}. Additional LAT detections using non-standard data types and techniques developed post-launch are also described by \citet{LAT_cat}. 

The GBM design is largely based on the Burst and Transient Source Experiment (BATSE) on the Compton Gamma Ray Observatory \citep{Fish89}, which operated from 1991 to 2000.  Both instruments employ multiple sodium iodide (NaI) detectors to achieve full sky field of view, have on-board burst triggering capability, and use relative count rates to obtain approximate directions to bursts. GBM also includes two bismuth germanate (BGO) detectors that can better detect higher energy photons. BATSE, with significantly larger NaI detectors, had better sensitivity, while GBM has a broader energy range and higher data rate.

This catalog summarizes some basic characteristics of the triggered GRBs: sky location, duration, peak flux and fluence. Spectral characteristics derived from a standard analysis are described in a companion catalog \citep{spec_cat}. Detailed studies of various GBM GRB subsamples have been presented elsewhere \citep{Gur10, Ghirl10, Lv10, Biss11, Gru11, Ghirl11, Nava11a, Nava11b, Zhang11}.

\section{Instrumentation}

GBM comprises twelve NaI scintillation detectors and two BGO scintillation detectors. The NaI detectors are 0.5~in.\ thick by 5~in.\ diameter and operate in the energy range 8~keV to 1~MeV. The performance of the NaI detectors at low energies is significantly enhanced by the use of beryllium entrance windows. Their positions and orientations on the spacecraft permit burst localization over the entire sky (unocculted  by the Earth). The BGO detectors are 5~in.\ thick by 5~in.\ diameter and operate in the 200~keV to 40~MeV energy range. They are located on opposite sides of the spacecraft so that at least one of them is illuminated from any direction. The GBM flight software (FSW) continuously monitors the detector count rates to detect GRBs and other short-timescale transients, computes their location on the sky, provides a preliminary classification, and promptly notifies the LAT of their occurrence. Once triggered, the FSW initiates prompt transmission of a subset of the data to the ground for quick-look analysis and notification of ground-based instruments via the Gamma-ray Coordinates Network (GCN). The instrument is described in more detail by \citet{Meegan}.

The GBM trigger algorithms operate on background-subtracted count rates over a programmable range of timescales (from a minimum of 16~ms to a maximum of 16.384~s; currently the longest is 4~s) and in four different energy ranges (currently 25--50~keV, 50--300 keV, $> 100$~keV and $> 300$~keV. This is the primary method for detection of GRBs, TGFs, SGRs and solar flares.

\Fermi's use of the Ku band for downlink of science data allows GBM to have a higher data rate than BATSE, which generally translates to better time and energy resolution. Outside of a trigger, the GBM continuously transmits two types of science data: continuous time (CTIME) and continuous spectroscopy (CSPEC). The CTIME data have finer time resolution (nominally 256~ms but configurable from 64~ms to 1.024~s in units of 64~ms) and coarse energy resolution (8~channels). The CSPEC data have the full energy resolution (128~channels) but more coarse time resolution (nominally 4.096~s but configurable from 1.024~s to 32.768~s in units of 1.024~s). In this mode time-tagged event (TTE) data are also produced but not transmitted to the ground. The TTE data consist of individual detector events, each tagged with arrival time (2~$\mu$s resolution, accurate to $\sim$10~$\mu$s), energy (128~channels) and detector number. These non-triggered TTE data are temporarily stored on-board in a ring buffer with a capacity of 512,000 events, which lasts for 25-30~s at typical background rates.

Upon entering trigger mode, the FSW speeds up CTIME resolution to 64~ms and CSPEC resolution to 1.024~s. In addition, the TTE data are transmitted directly to the science data bus instead of being stored in the ring buffer (the contents of which are frozen for later downlink). The production of prompt TTE lists for approximately 300~s, at which point the contents of the pre-trigger ring buffer are dumped to the science data bus. After an additional 300~s, the CTIME and CSPEC data are returned to their non-triggered time resolution and the FSW exits trigger mode. 

The basic trigger design follows that used for BATSE: to trigger, two or more detectors must have a statistically significant rate increase above the background rate. Requiring two detectors to be above their respective thresholds makes triggering on statistical fluctuations less likely, and much more importantly, it suppresses triggering due to non-astrophysical effects that appear in only one detector, such as phosphorescence spikes. Each algorithm has its own threshold setting, configurable from 0.1~$\sigma$ to 25.5~$\sigma$ in units of 0.1~$\sigma$. The background model is a trailing average of the detector data. Whereas BATSE used three trigger algorithms (a single energy range, usually 50--300~keV, and three timescales, 64~ms, 256~ms \& 1.024~s), the GBM FSW supports up to 119 trigger algorithms. A given algorithm is defined by its timescale, offset and energy range. The offset is a value in milliseconds by which the time binning is shifted. Running overlapping accumulations for a given combination of timescale and energy range provides some improvement in trigger sensitivity \citep{Band02,Band04}.

While in trigger mode, the FSW continues to monitor other enabled algorithms and records all instances where a given algorithm would also have triggered on the same burst. A special \textit{compute} mode is available in which an algorithm is monitored during a trigger (initiated by any other algorithm) to determine if its threshold is exceeded but no triggers can actually be initiated by that algorithm.

The FSW classifies triggers based on a number of criteria, including the event localization, spectral hardness, and the spacecraft geomagnetic latitude (McIlwain L coordinate). A Bayesian approach is used to assign identification probabilities for various event classes, including GRB, solar flare, SGR, particle precipitation and known transient sources. Classification of TGFs is a special case: early in the mission it was observed that these events trigger NaI detectors exclusively on the combination of shortest timescale (0.016~s) and one of the high energy ranges ($> 100$~keV or $> 300$~keV).

While in trigger mode, the FSW continues to monitor the detector rates on various timescales and, based on improved statistical significance, updates of the localization and classification may be generated and transmitted in the quick-look science data.

\section{In-Orbit Operations}

\subsection{Trigger Criteria}

\label{trigger_criteria}
GBM triggering has been enabled continuously since 12 July 2008, except during South Atlantic Anomaly passages and a few brief intervals when FSW upgrades were being installed.
Early in the mission the trigger algorithms used only data from the NaI detectors. This turned out to be a limitation for TGFs because they are spectrally much harder than GRBs. In November 2009 the flight software was revised to add trigger algorithms that use the BGO detectors, which significantly improved the GBM sensitivity for detecting TGFs.

In orbit, the GBM has enabled 71 trigger algorithms, five of which are TGF-specific algorithms that use the BGO detectors alone or combined with NaI detectors. No GRB has ever triggered only on a TGF-specific algorithm, so they are not discussed further in this paper. Table \ref{trigger:criteria:history} summarizes the 66 algorithms relevant for this catalog, the times during which they were enabled and the history of their threshold settings.

During the first year most of the algorithms were enabled and a few minor adjustments made to their thresholds. Exceptions are the 16-second algorithms (numbers 20, 21, 41 \& 42), which were deemed too sensitive to background variations and disabled after $\sim$2~weeks. After nearly a year's experience, it was judged that a large number of the algorithms were of dubious value because they never triggered on an event that did not also trigger another algorithm. This list included all of the 8-second algorithms as well as most of the algorithms not operating in the standard BATSE energy range (50--300~keV). In order to ease the computational burden on the FSW, these algorithms were disabled in early July of 2009. The configuration in the rightmost column of Table \ref{trigger:criteria:history} remained the same through the remainder of the period covered by this catalog.

Direct comparisons between the BATSE and GBM datasets are possible since GBM includes the same three trigger algorithms used by BATSE (64, 256, and 1024 ms time scales in the 50 to 300 keV energy range). Depending on which set of the half-bin offset algorithms are considered as the BATSE algorithms, we find that either 405 or 408 of the 491 GRBs would have triggered GBM, in agreement with pre-launch estimates of ~200 bursts/year \citep{Meegan07}. Using both sets of overlapping windows raises the total number of triggers to 423, leaving 68 events which did not trigger on any BATSE-style algorithms. Of these, 63 (93\%) triggered the longer ($> 1.024$~s) timescale algorithms in the 47 to 291 keV energy range, three triggered only the 512~ms algorithm in 47-291 keV, and two triggered only in the 23-47 keV energy range (one each in 64~ms and 2.048~s). Thus, the apparent improvement in trigger sensitivity relative to BATSE is attributable mainly to GBM's additional longer trigger timescales.

\subsection{Quick-look Analysis}
\label{QLA}

As described above, quick-look data are generated during trigger mode and promptly transmitted to the ground. For events classified as GRBs, the FSW-generated localization and classification information is further distributed via GCN notices. Also for GRBs, additional GCN notices containing ground-generated localizations are produced and distributed automatically.  The GBM location algorithm is an adaptation of the method developed for BATSE \citep{LOCBURST}. Both FSW and ground locations use the count rates in all 12 NaI detectors to point back to a preferred direction on the sky by comparing observed rates to model rates and minimizing $\chi^2$. The model rates are a combination of counts that come directly from the source into the detector, counts from the source scattered in the spacecraft into the detectors, and counts from source photons that hit the Earth's atmosphere and are scattered into the detectors.  All three of these components are a function of the source intensity, its spectrum, and the source-spacecraft geometry,  with the final component also depending on the source-spacecraft-Earth geometry.   For automated locations onboard and on the ground, the background count rate subtracted from the observed counts is an average over a 16~s interval before the burst trigger occurred. However, the ground automated localizations differ from the flight locations in several ways:
\begin{enumerate}
\item Although the two decision making processes use the same rates data type, they run independently with different criteria and do not necessarily use rate data from the same time intervals.
\item  The ground process has access to location tables generated with finer sky resolution (1\degr, compared to 5\degr\ for the FSW).
\item The ground process includes a more accurate treatment of atmospheric scattering (based on the actual orientation of the spacecraft with respect to the Earth, whereas the FSW assumes zenith-pointing for all model rates).
\item The ground process incorporates the spectrum of the source into the calculation of the expected rates by choosing one of three location tables based on the hardness of the burst as determined by the ratio of counts $> 50$~keV to counts $< 50$~keV.
\end{enumerate}
 
The GBM team assigns a burst advocate (BA) to inspect the real-time data promptly and perform additional analysis as appropriate. Normally the BA will generate additional localizations and optionally distribute these via the GCN (circulars were used during the time period of this catalog but currently GCN notices are used). These ``human-in-the-loop'' localizations use source and background time intervals and model fits selected by the user based on the the entire quick-look data set, which extends from 200~s pre-trigger to 450~s post-trigger. 
The BAÕs typically run the location code several times, using different selections of time interval and/or background models, and select a ÒbestÓ location using statistical error and goodness-of-fit criteria. This is particularly useful in verifying that separate pulses are consistent with the same sky location. 
The FSW classification is reviewed by the BA, usually in consultation with other GBM team members, and may be corrected based on inspection of the GBM quick-look data and/or additional information such as detection by another instrument. 

\section{Catalog Analysis}

\subsection{Burst Localization and Instrument Response}


Determination of the approximate burst sky location is important because the other results reported in this catalog and the companion spectroscopy catalog \citep{spec_cat} require instrument response functions that are dependent on the direction of the burst relative to the detectors and to the Earth. Most of the burst locations reported in this catalog are the result of the manual on-ground analysis, typically by the BA as described in Section~\ref{QLA}. If, however, the burst was also localized with better precision by another instrument (e.\ g., \textit{Swift} or the LAT) that location was used instead to derive the GBM instrument response for the subsequent catalog analysis. A total of 76 bursts have locations from \textit{Swift}; 63 of these triggered the \textit{Swift} Burst Alert Telescope (BAT) and 5 more were detected in ground analysis of BAT data. The remaining 8 were located by the \textit{Swift} X-Ray Telescope following detections of prompt emission by other instruments. 

The accuracy of the GBM burst localizations was checked by comparing the independently-derived GBM locations with a sample of higher precision locations obtained by other instruments for the same GRBs.
Using 127 bursts with known locations (some of which occurred after the end of the current catalog), we find that the true GBM human-in-the-loop location errors are best described by combining the statistical error in quadrature with a systematic error, where the current best model for systematic errors is 2.6\degr\ with 72\% weight and 10.4\degr\ with 28\% weight. As the actual statistical error contours are not circular, it is instructive to see how many of the more precise burst locations fall within our quoted statistical error circle. Of those 127 bursts, 51 (40\%) are within the 1~$\sigma$ statistical error radius, 93 (73\%) are within twice the 1~$\sigma$ radius and 107 (84\%) are within 3 times the 1~$\sigma$ radius.
A more detailed analysis of the GBM location errors is in progress and will be reported later \citep{Briggs11}. 

\begin{figure}
\begin{center}
\plotone{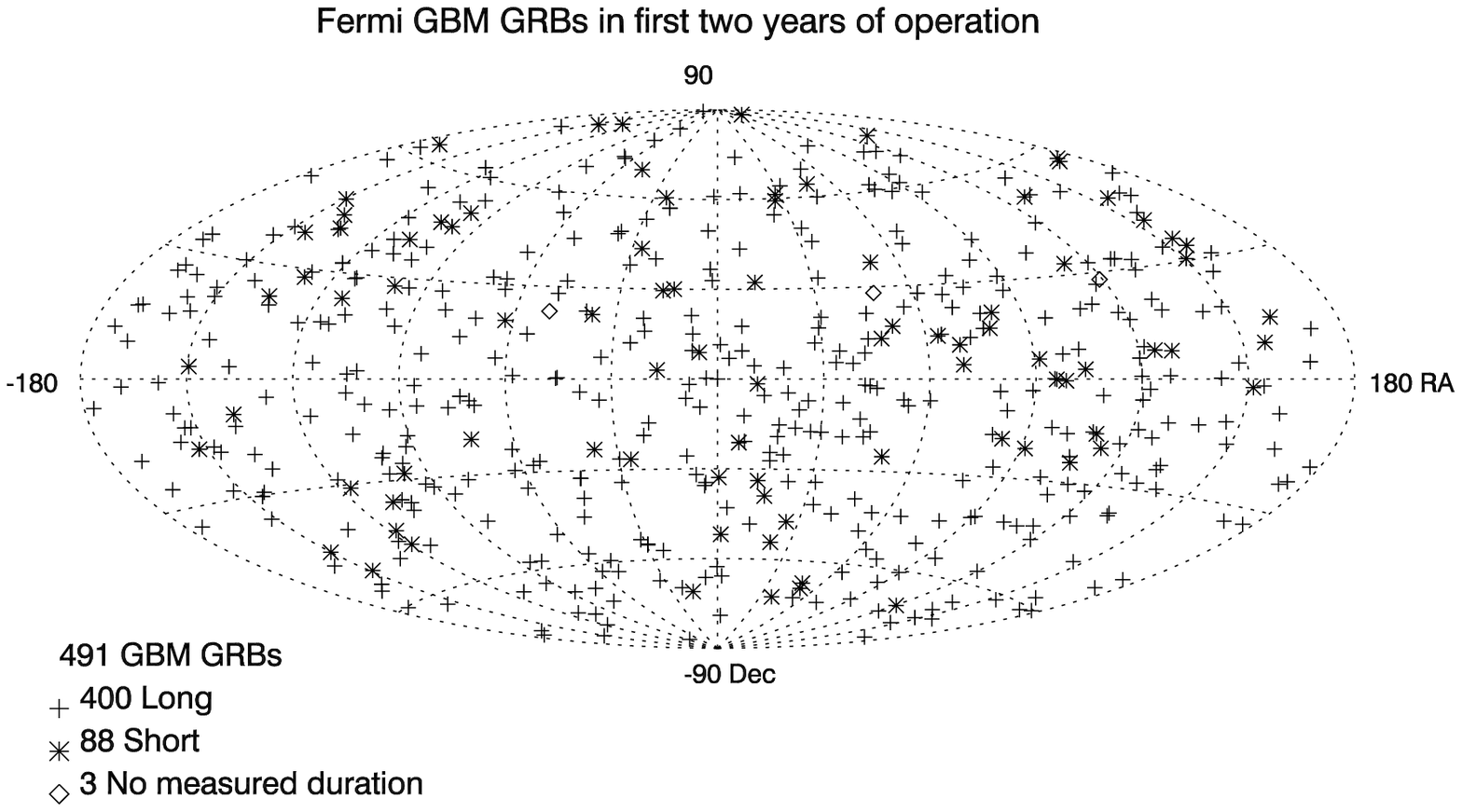}
\caption{\label{sky_dist} Sky distribution of GBM triggered GRBs in celestial coordinates. Crosses indicate long GRBs ($T_{90} > 2$~s); asterisks indicate short GRBs.}
\end{center}
\end{figure}

Figure~\ref{sky_dist} shows the sky distribution of the GBM-detected GRBs in celestial coordinates.  The large-scale isotropic distribution is well-known from BATSE observations \citep{Briggs96} and the GBM distribution appears to be consistent with this. 

\subsection{Duration, Peak Flux \& Fluence}

In addition to the burst locations, the present paper reports various measures of the duration, peak flux and fluence of each burst, with a few exceptions due to analysis difficulties such as incomplete data or background interference. The burst durations $T_{50}$ and $T_{90}$\footnote{$T_{50}$ is the interval between the times 
where the burst has reached 25\% and 75\% of its fluence. $T_{90}$ is similarly defined between 5\% and 95\% of the fluence.} were computed in the 50-300~keV energy range. The fluence for each burst was computed in two energy ranges: 50-300~keV and 10-1000~keV. Peak fluxes for each burst were computed in these same energy ranges and for three different timescales: 64~ms, 256~ms and 1024~ms.

Burst durations were determined using a method similar to that developed for BATSE \citep{CK93, Koshut}. However, in the BATSE analysis all quantities were derived from the counts directly, whereas in the present analysis the counts spectrum in each time bin is deconvolved and the durations are computed from the time history of fitted photon spectra. Peak fluxes and fluences are naturally obtained in the same analysis, using the same choices of detector subset, source and background intervals and background model fits. 
Unlike CGRO, which was inertially pointed in the same orientation for weeks at a time, the \Fermi\ observatory has been operated in an all-sky survey mode during the period covered by this catalog. To optimize sky coverage the spacecraft sweeps its z-axis across the sky at a specified angle perpendicular to the orbit plane (currently 50\degr), rocking on alternate orbits above and below the orbit plane by the specified angle. Within a given orbit \Fermi\ also executes a slow roll about the z-axis to maintain optimal orientation of the solar panels with respect to the sun.
In the energy range of interest the response of a given GBM NaI detector varies approximately as the cosine of the angle between the source direction and the detector axis. Therefore as Fermi slews, the detector to source angle changes and hence so does the response, with the rate of change being different for each detector. This was not a factor for BATSE, where inertial pointing kept the source to detector angles constant for extended periods. 
The most accurate correction for the response changing over time is a spectral deconvolution (assuming that the changing 
response is tracked correctly), so we have adopted the deconvolved flux history as our basis for determining the duration parameters. 
Furthermore, comparisons from burst to burst of the peak flux and fluence will not be compromised by differences in response arising from 
different source angles, as they would be for the raw counts. Finally, a fluence that is obtained by integrating a deconvolved flux history 
incorporates spectral evolution throughout the event in a way that summed counts can never attain, due to the loss of temporal information 
from the summing. 
Appendix~A describes the procedure in some detail.


For each burst, a set of NaI detectors was chosen with good source viewing angles ($ < 60$\degr) and no apparent blockage by any other element of the spacecraft. For the majority of bursts the GBM CTIME data, which have 256~ms time resolution pre-trigger and 64~ms resolution post-trigger, were used. TTE data were used for bursts where at least one of the peak fluxes occurs at or before the trigger time, which happens for many short bursts and a few longer ones. A limitation is that the pre-trigger TTE data typically span at most 30~s, which in some cases was not enough for computing the background and for some long bursts included significant burst emission. In such cases, the analysis was done with CTIME data. When using TTE data, which have 256 channels of energy resolution, it was often found that the deconvolution analysis is more robust if the 256 pulse-height channels were first summed into 8 channels, as in the CTIME data. Because of the relatively small number of bursts with detectable emission in the BGO detectors, only data from the NaI detectors was used for the catalog analysis.

\section{Catalog Results}

The catalog results can be accessed electronically through the HEASARC browse interface (http://heasarc.gsfc.nasa.gov/W3Browse/fermi/fermigbrst.html). Standard light curve plots for each burst can be viewed at http://gammaray.nsstc.nasa.gov/gbm/science/grbs/month\_listings/. Here we provide tables that summarize selected parameters.

Table~\ref{main_table} lists the 492 triggers that were classified as GRBs. The GBM Trigger ID is shown along with a conventional GRB name as defined by the GRB-observing community. 
For readers interested in the bursts with significant emission in the BGOs, the trigger ID and GRB name are highlighted in italics if emission in the BGO data (above 300 keV) is visible in the standard light curve plots.\footnote{These BGO-detected identifications are the result of a visual search rather than a quantitative analysis and thus do not have a well-defined threshold.}
 Note that the entire table is consistent with the small change in the GRB naming convention that became effective on 1-Jan-2010 \citep{Barth09}: if for a given date no burst has been ``published'' previously, the first burst of the day observed by GBM includes the 'A' designation even if it is the only one for that day. 
The table lists the GBM-derived location only if no higher-accuracy locations have been reported by another instrument. The choice of a higher-accuracy location is somewhat arbitrary (e.\,g., Swift-BAT locations are often listed even if a Swift-XRT location is available); for the GBM analysis, location accuracy better than a few tenths of a degree provides no added benefit. The table also shows which algorithm was triggered along with its timescale and energy range. Note that the listed algorithm is the first one to exceed its threshold but it may not be the only one. Finally, the table lists other instruments that detected the same GRB. Many of these are determined by inspection of web-based tables and/or light curves. For some instruments (e.\,g., Suzaku-WAM) no automated trigger was generated but the GRB is clearly visible in a web-accessible light curve. Those cases are shown with an asterisk in Table~\ref{main_table}.

The results of the duration analysis are shown in Tables~\ref{durations}, \ref{pf_fluence} \& \ref{pf_fluence_b}. The values of $T_{50}$ and $T_{90}$ in the 50--300~keV energy range are listed in Table~\ref{durations} along with their respective statistical error estimates and start times relative to the trigger time.  For a few GRBs the duration analysis could not be performed, due either to the weakness of the event or to technical problems with the input data. Also, for some GRBs the results are underestimates, either because of Earth-occultation or because the input data were truncated by SAA entry. Finally, for technical reasons it was not possible to do a single analysis of the unusually long GRB 091024A \citep{091024}, so the analysis was done separately for the two triggered episodes. These cases are all noted in the Table. The reader should also be aware that for most GRBs the analysis used data binned no finer than 64~ms, so the duration estimates (but not the errors) are quantized in units of 64~ms. For a few extremely short events (noted in the table) TTE data were used with 32~ms or 16~ms binning. 

As part of the duration analysis, peak fluxes and fluences were computed in two different energy ranges. Table~\ref{pf_fluence} shows the values in 10-1000~keV and Table~\ref{pf_fluence_b} shows the values in 50--300~keV. 
As discussed in Appendix~A, the analysis results for low fluence events are subject to large systematic errors and should be used with caution.

\section{Discussion}

\begin{figure}
\begin{center}
\plotone{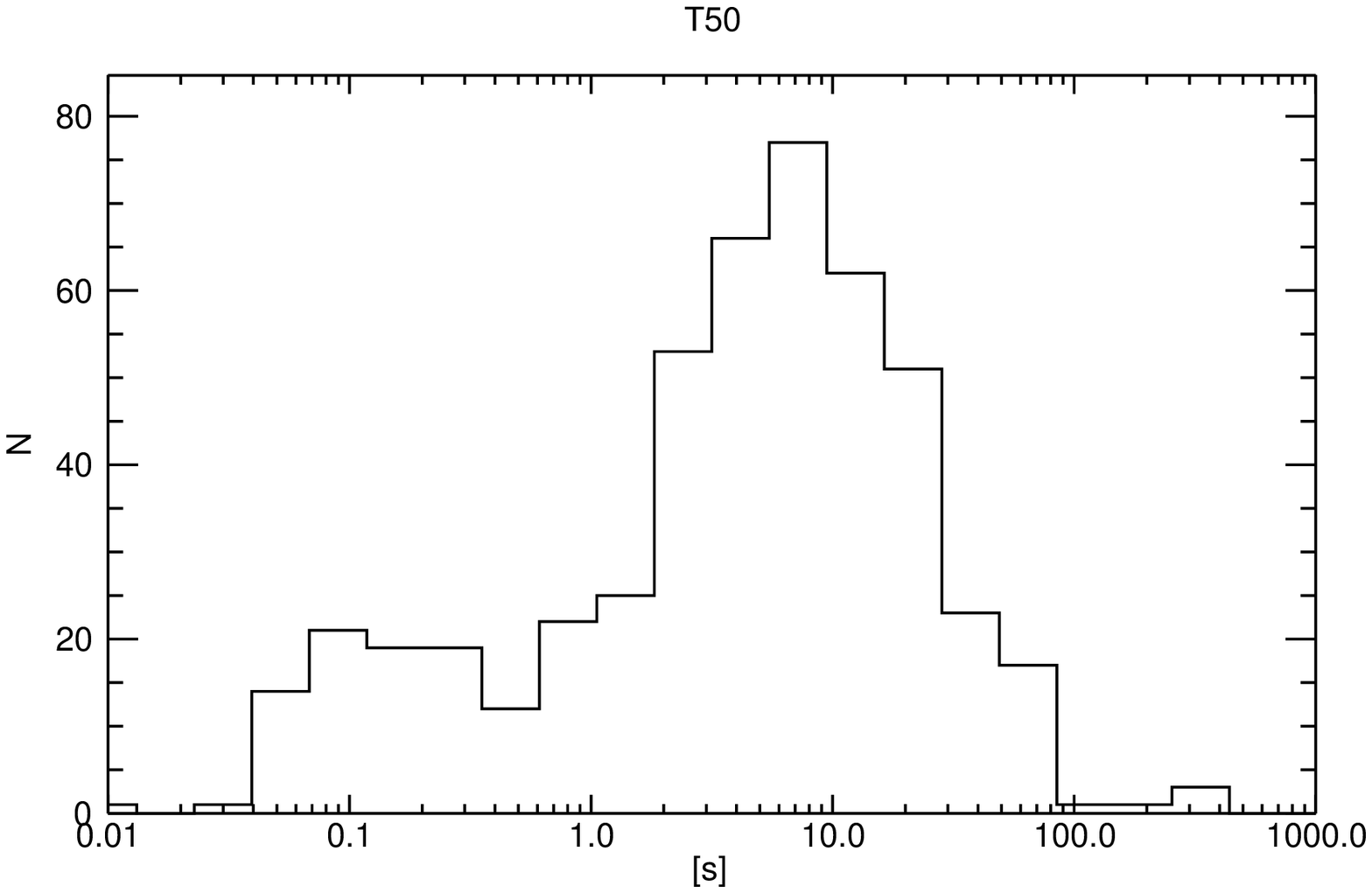}
\plotone{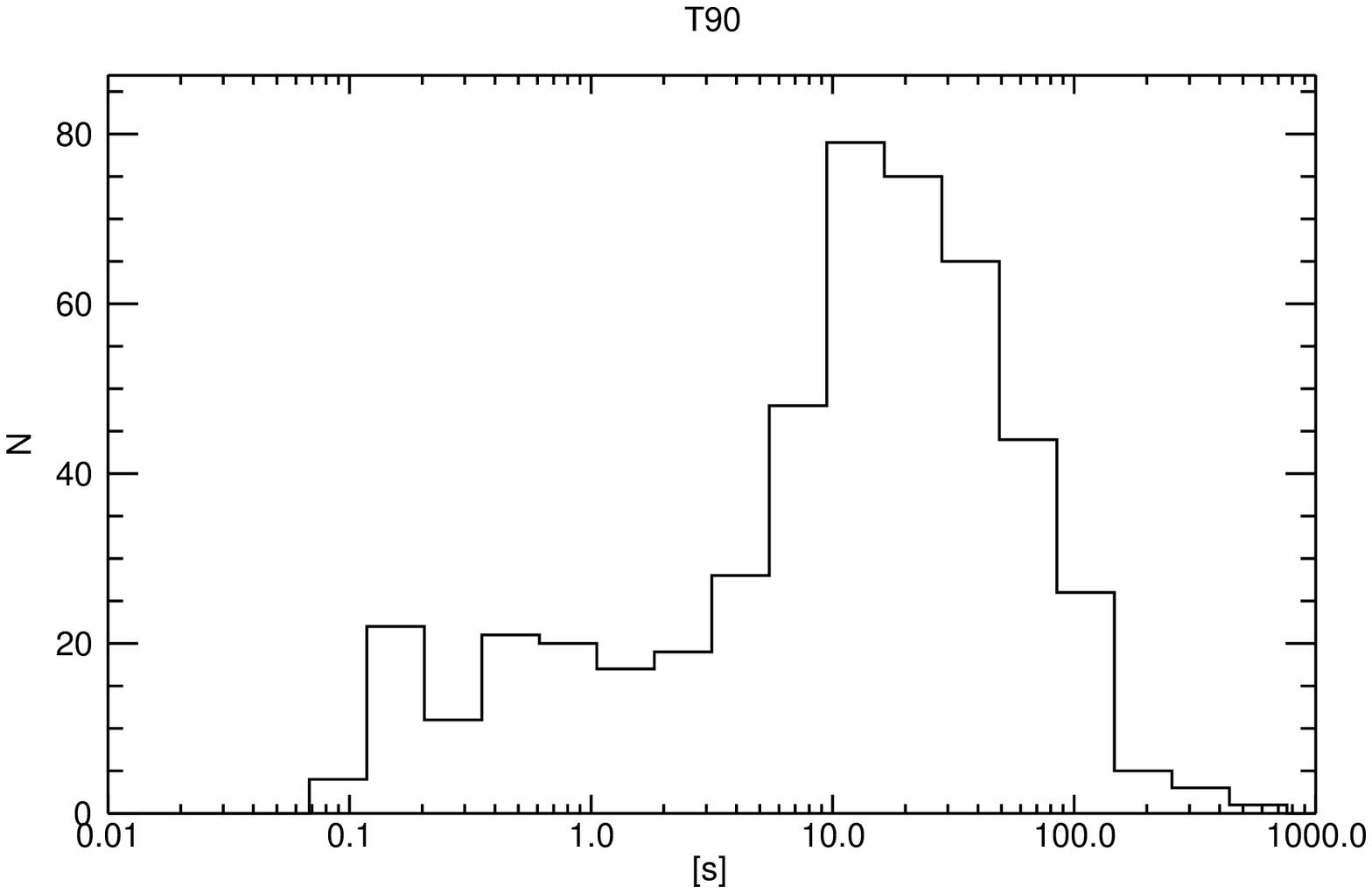}
 \caption{\label{dur_dist} Distribution of GRB durations in the 50--300~keV energy range. The upper plot shows $T_{50}$ and the lower plot shows $T_{90}$. }
\end{center}
 \end{figure}

Histograms of the $T_{50}$ and $T_{90}$ distributions are shown in Figure~\ref{dur_dist}. Using the conventional division between the short and long GRB classes ($T_{90} = 2$~s), we find 88 (18\%) of the 487 measured GRBs to be in the short classification. Within the quoted duration errors, the number of short GBM events ranges from 73 (15\%) to 104 (21\%). For comparison, the fraction of short events in the BATSE GRB catalog is 24\%. The difference from BATSE is probably not due to a deficit of short events but rather to an excess of long events detected by GBM's longer timescale trigger algorithms (see Section \ref{trigger_criteria}).

\begin{figure}
\begin{center}
\epsscale{0.9}
\plotone{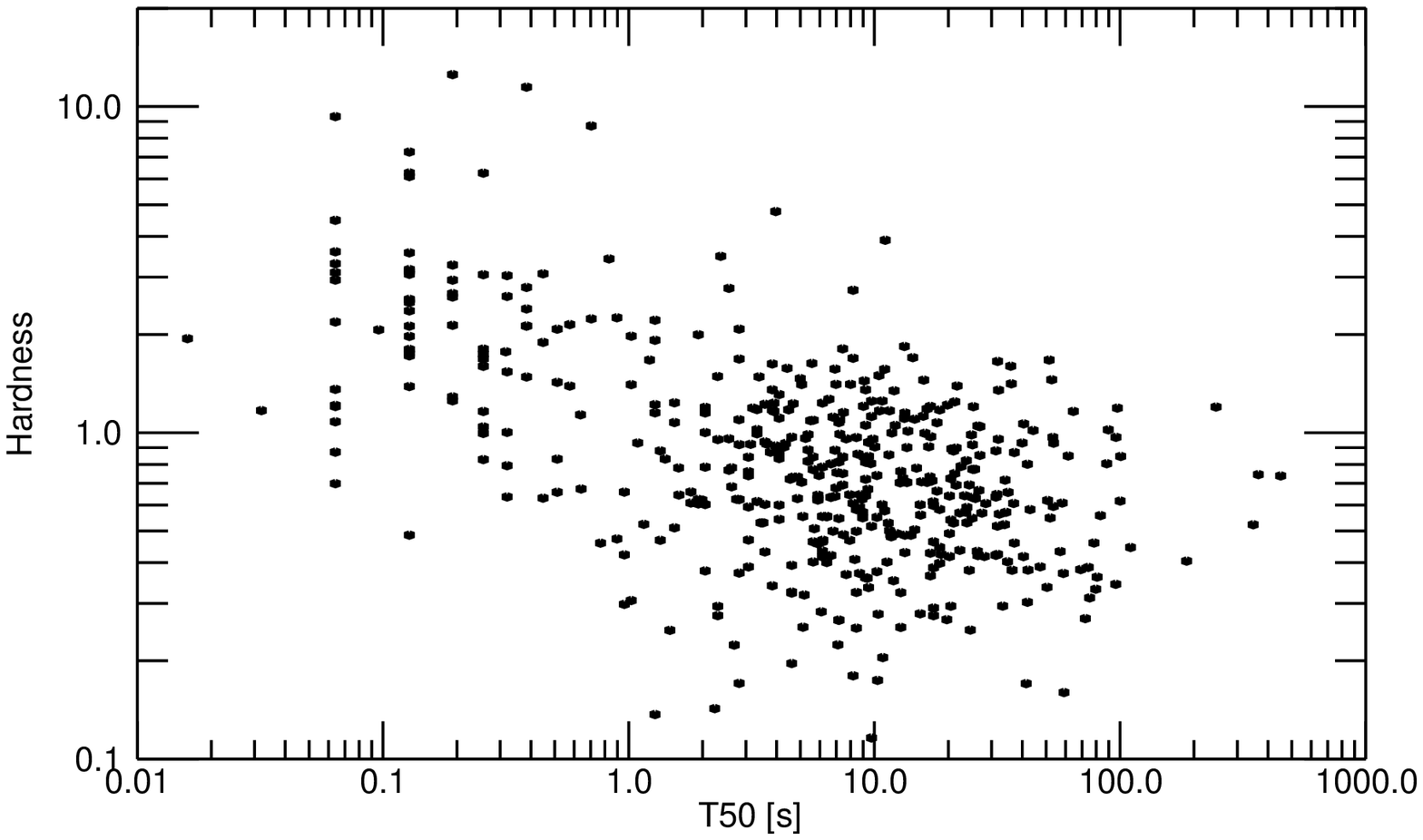}
\plotone{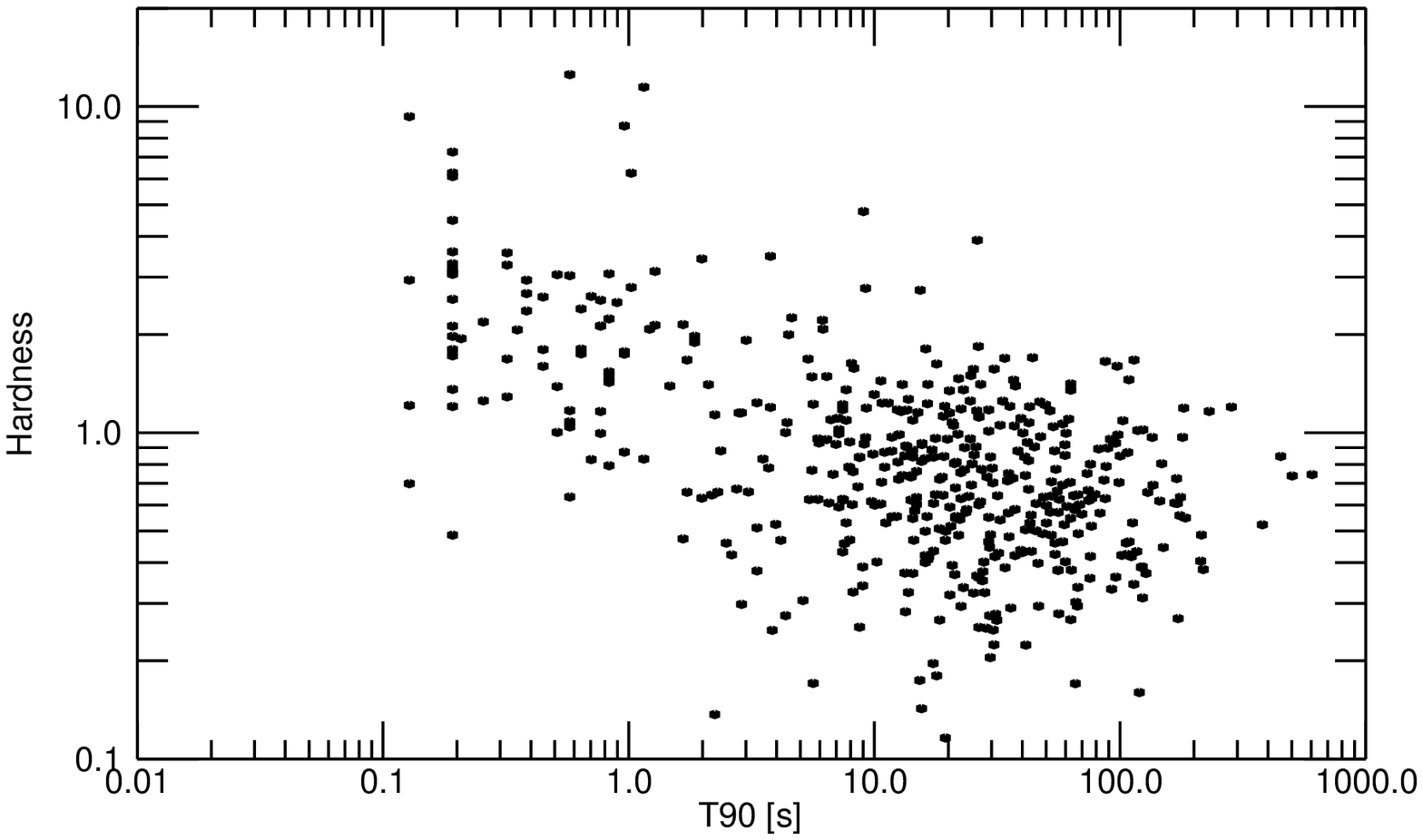}
 \caption{\label{hardness_vs_dur} Scatter plots of spectral hardness vs.\ duration are shown for the two duration measures $T_{50}$ (upper plot) and $T_{90}$ (lower plot). The spectral hardness was obtained from the duration analysis results by summing the deconvolved counts in each detector and time bin in two energy bands (10-50 and 50-300 keV), and further summing each quantity
in time over the $T_{50}$ and $T_{90}$ intervals.  The
hardness was calculated separately for each detector as the ratio of the flux density
in 50-300 keV to that in 10-50 keV and finally averaged over detectors. For clarity, the estimated errors are not shown but can be quite large for the weak events. Nevertheless, the anti-correlation of spectral hardness with burst duration is evident.} 
\end{center} 
\end{figure}

The anti-correlation of spectral hardness with duration is well known from BATSE data \citep{CK93} and a simple analysis shows that such an effect is also present in the GBM catalog. Time-resolved spectral fits for each GRB are a by-product of the duration analysis and
those photon model fit parameters were used to derive a measure of average spectral hardness. Figure~\ref{hardness_vs_dur} shows scatter plots of hardness derived in this way as a function of duration for the two duration measures. Although the effect of the 64-ms duration quantization is noticeable, the anti-correlation of hardness vs.\ duration is visibly evident in the GBM data.

\begin{figure}
\begin{center}
 \epsscale{0.9}
\plotone{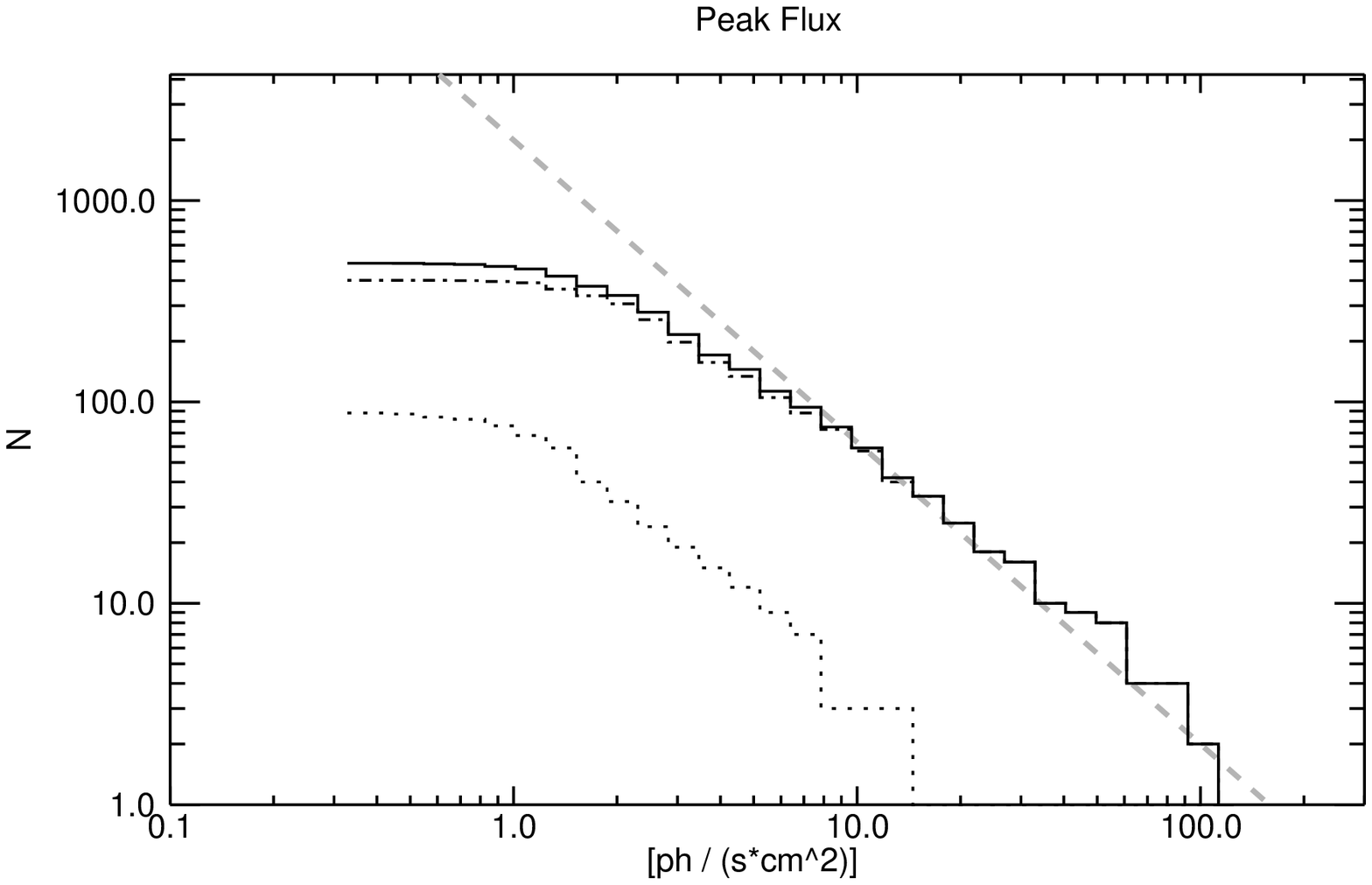}
\plotone{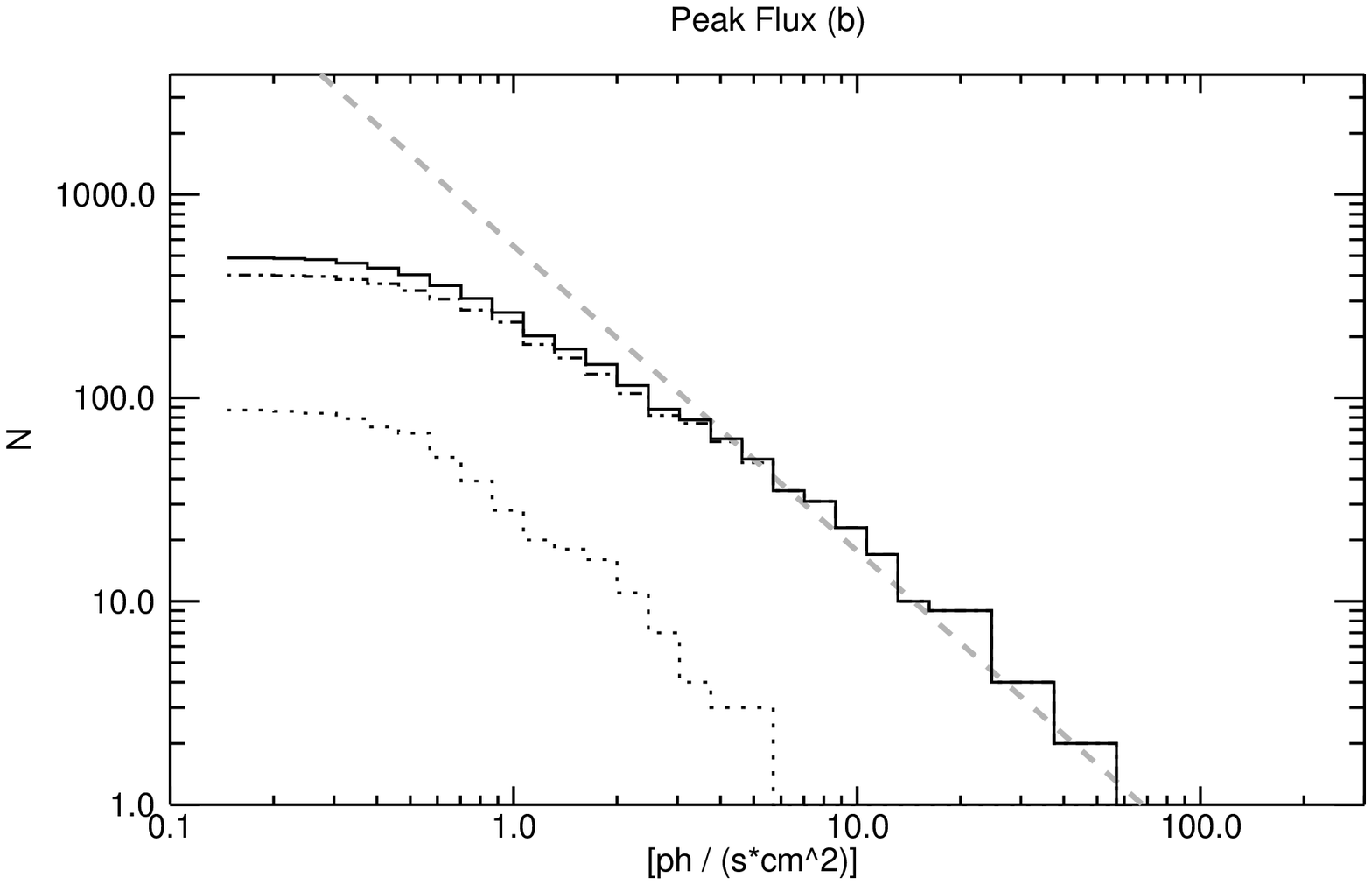}
\caption{\label{pflx_fig} Integral distribution of GRB peak flux on the 1.024~s timescale. Energy ranges are 10--1000~keV (upper plot) and 50--300~keV (lower plot). Distributions are shown for the total sample (solid histogram), short GRBs (dots) and long GRBs (dash-dots), using $T_{90} = 2$~s as the distinguishing criterion. In each plot a power law with a slope of $-3 / 2$ (dashed line) is drawn to guide the eye.}  
\end{center}
\end{figure}

\begin{figure}
\begin{center}
 \epsscale{0.9}
\plotone{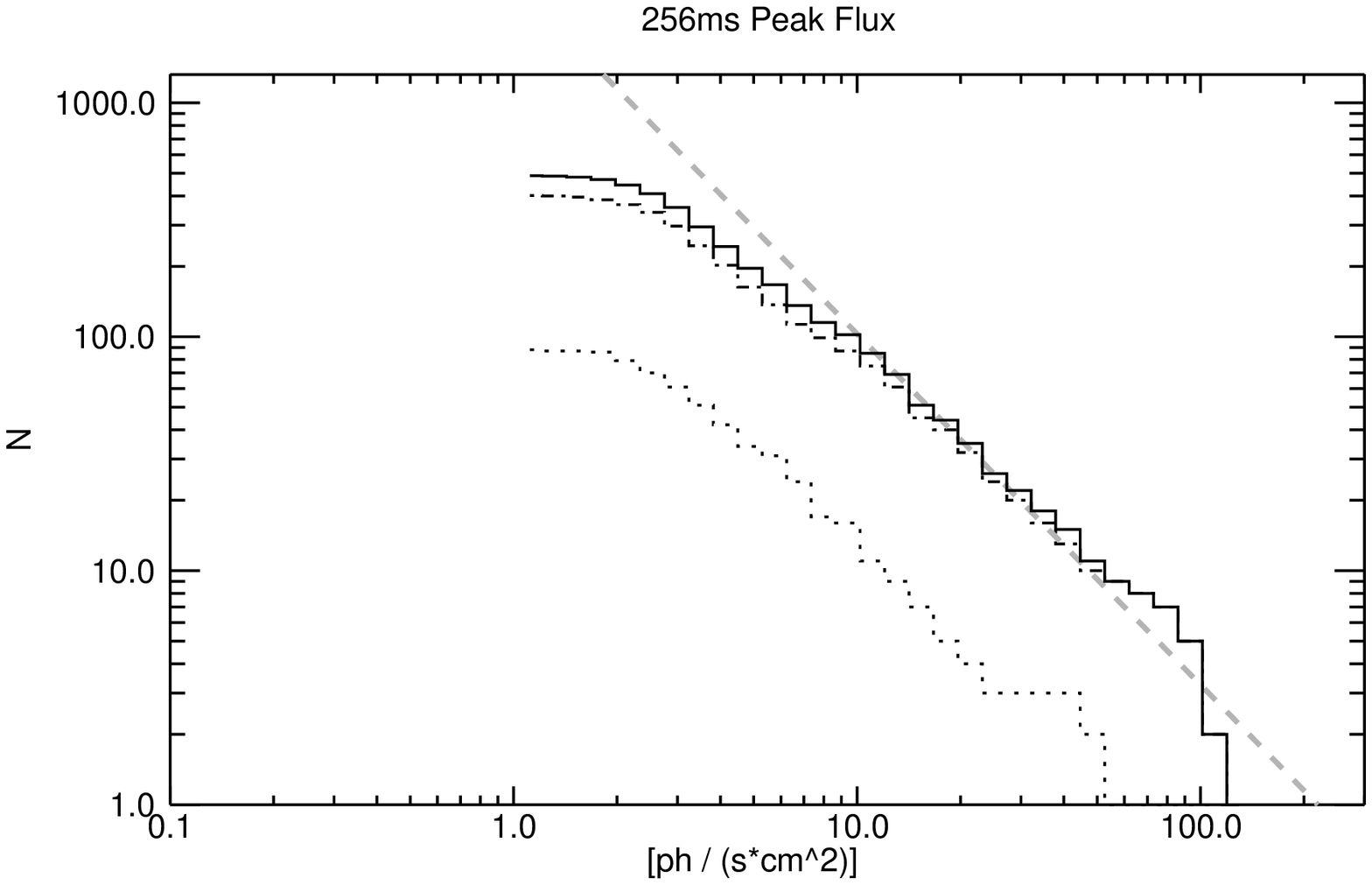}
\plotone{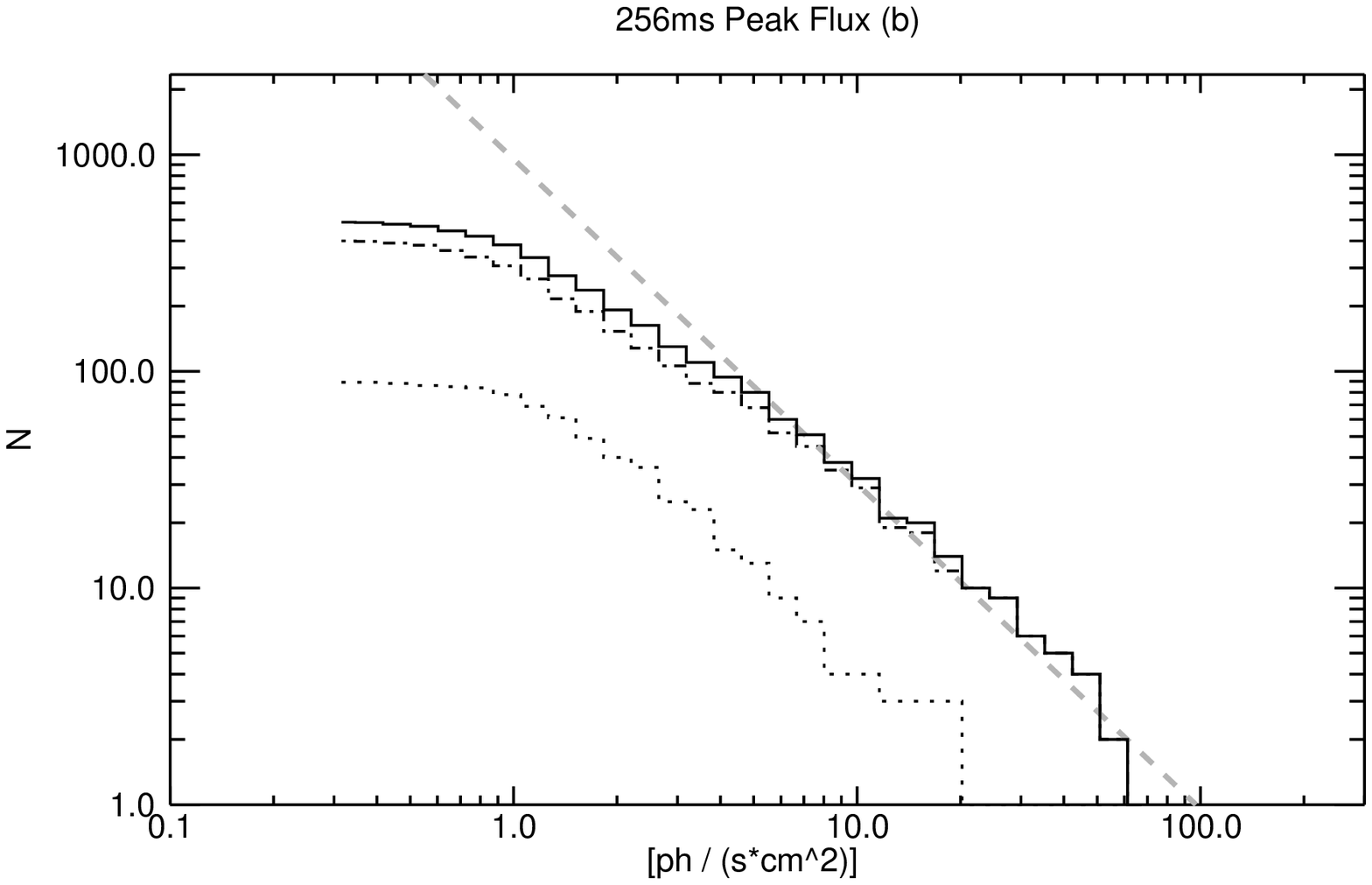}
\caption{\label{pf256_fig} Same as Figure~\ref{pflx_fig}, except on the 0.256~s timescale.} 
\end{center}
\end{figure}

\begin{figure}
\begin{center}
 \epsscale{0.9}
\plotone{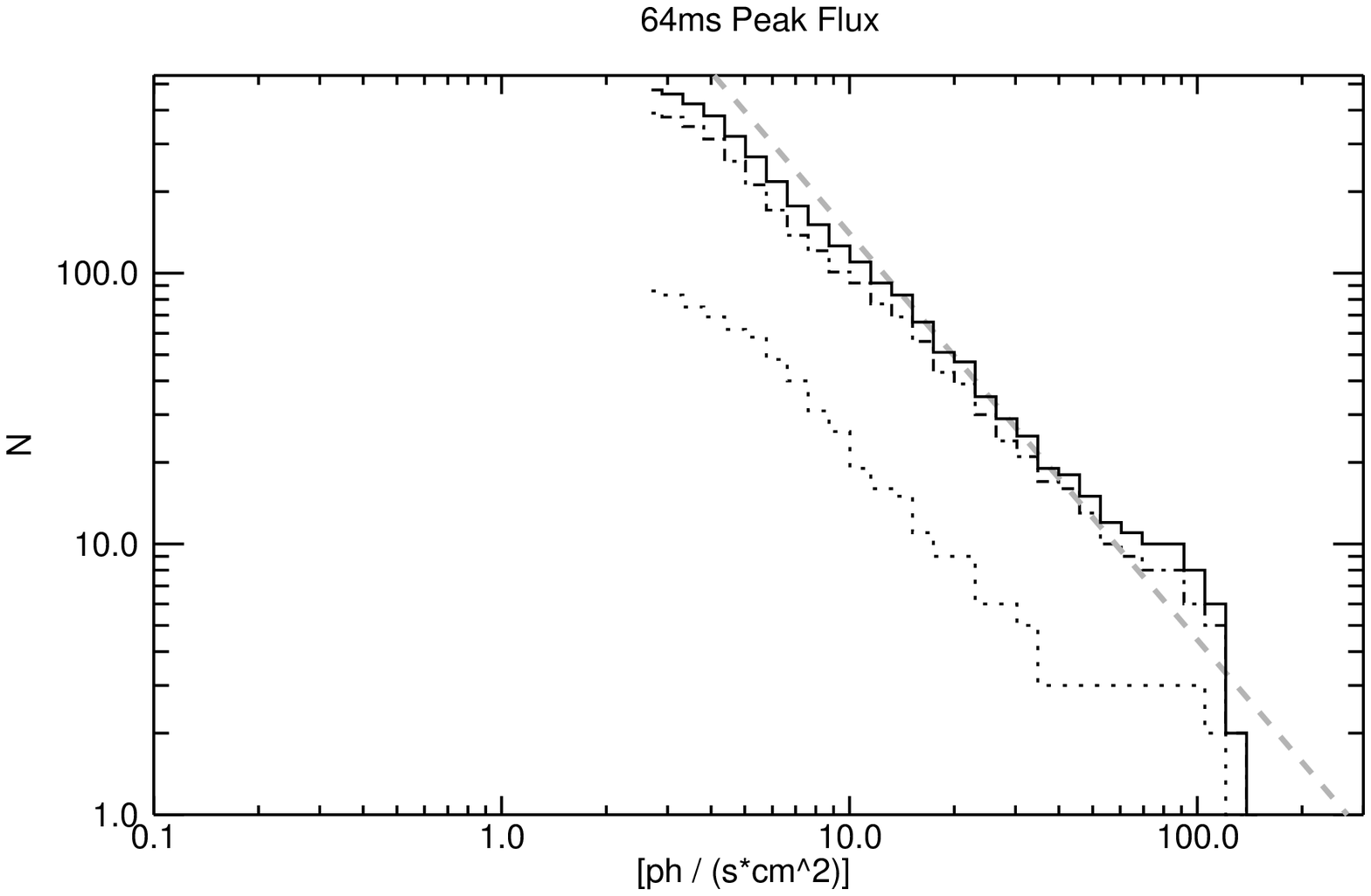}
\plotone{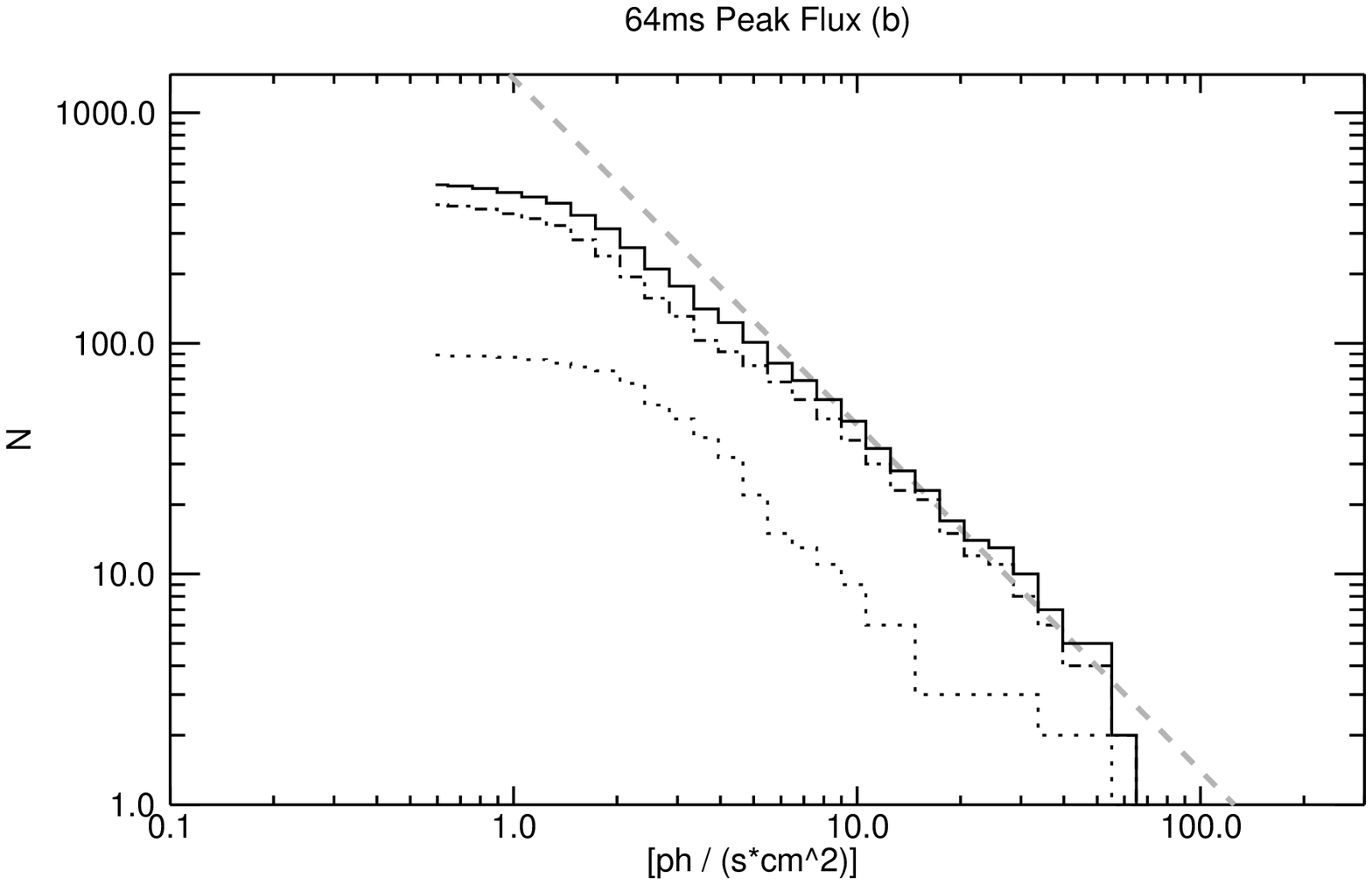}
\caption{\label{pf64_fig}Same as Figure~\ref{pflx_fig}, except on the 0.064~s timescale.} 
\end{center}
\end{figure}

\begin{figure}
\begin{center}
 \epsscale{0.8}
\plotone{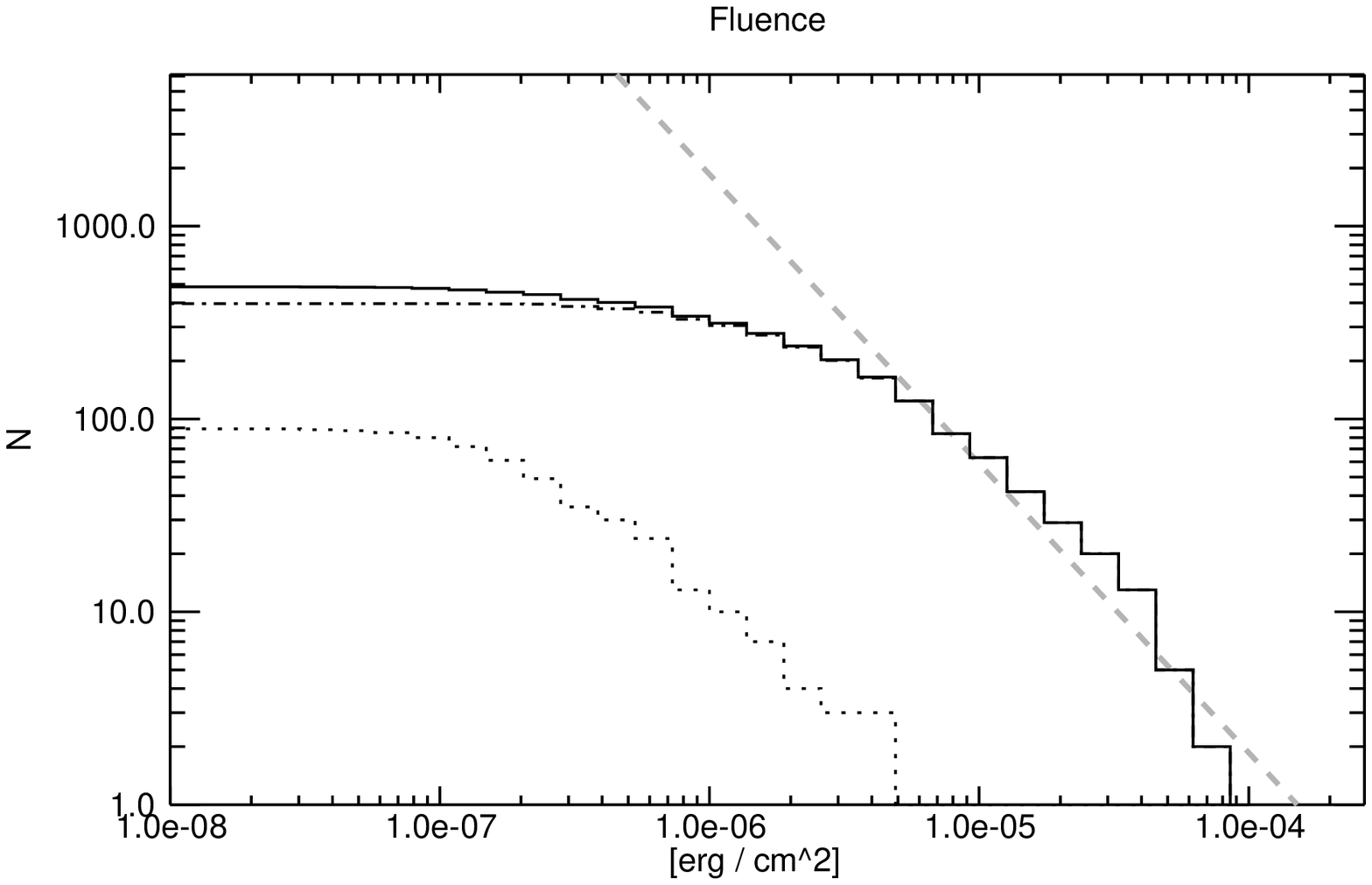}
\plotone{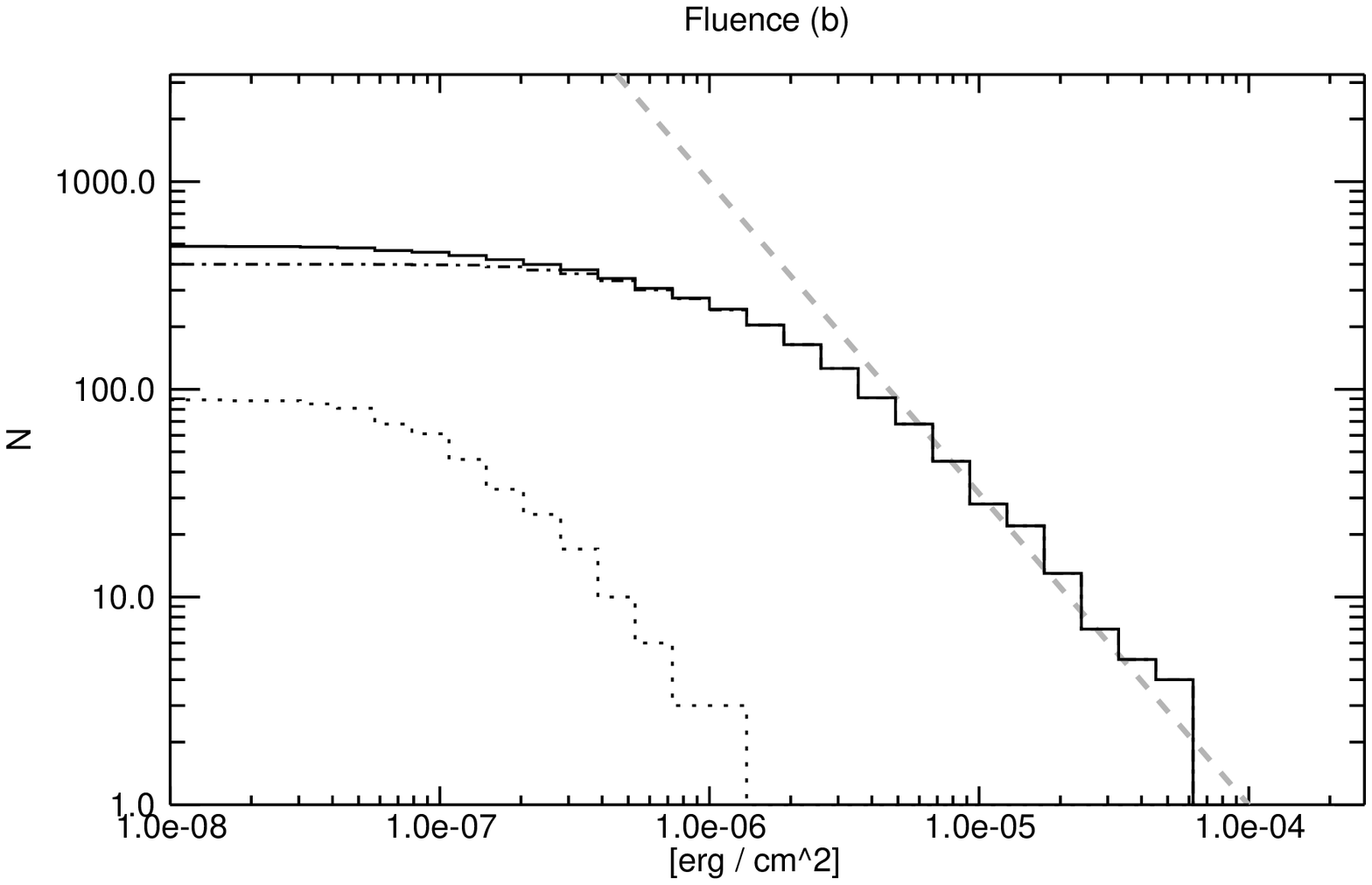}
\caption{\label{flu_fig} Integral distribution of GRB fluence in two energy ranges: 10--1000~keV (upper plot) and 50--300 keV (lower plot). Distributions are shown for the total sample (solid histogram), short GRBs (dots) and long GRBs (dash-dots), using $T_{90} = 2$~s as the distinguishing criterion. In each plot a power law with a slope of $-3 / 2$ (dashed line) is drawn to guide the eye.} 
\end{center}
\end{figure}

Integral distributions of the peak fluxes are shown in Figures~\ref{pflx_fig}--\ref{pf64_fig} for the three different timescales and separately for short and long GRBs. For the long GRBs, deviation from the $-3/2$ power-law that would be expected if the GRBs were spatially homogeneous occurs well above the GBM threshold at a flux value of $\sim$10 ph s$^{-1}$ cm$^{-2}$. This is consistent with earlier BATSE measurements \citep{Pacie99}, which have much better statistics. For the short events the GBM data appear consistent with a homogeneous spatial distribution down to peak flux values around 1~ph s$^{-1}$ cm$^{-2}$ (50-300 keV), below which instrument threshold effects become dominant. The integral fluence distributions for the two energy intervals are shown in Figure~\ref{flu_fig}.


\section{Summary}

The first GBM catalog includes 491 cosmic gamma-ray bursts that triggered GBM between 12 July 2008 and 11 July 2010. Compared to BATSE, GBM has a higher threshold for burst detection but this is somewhat ameliorated by GBM's additional range of trigger timescales (primarily the 4~s timescale) and, to a lesser extent, trigger energy ranges. The distribution of GBM durations is consistent with the well-known bimodality measured previously. The fraction of short GRBs in the GBM sample is somewhat smaller than detected by BATSE, which is attributed mainly to GBM's ability to trigger on longer timescales.

\acknowledgments

The GBM project is supported by NASA and by the German Bundesministeriums f\"ur Wirtschaft und Technologie (BMWi)  via the Deutsches Zentrum f\"ur LuftÐ und
Raumfahrt (DLR) under the contract numbers 50 QV 0301 and 50 OG 0502.
AG acknowledges the support of the Graduate Student Researchers Program funded
by NASA. SMB acknowledges support of the European Union Marie Curie Reintegration
Grant within the 7th Program under contract number PERG04-GA-2008-239176. SF
acknowledges the support of the Irish Research Council for Science, Engineering, and
Technology, co-funded by Marie Curie Actions under FP7.

\appendix

\section{Computation of Duration, Peak Flux and Fluence}

For the catalog analysis, a standard calculation of durations, peak fluxes and fluences was implemented as an add-on to the RMFIT software package that was developed for time-resolved analysis of BATSE GRB data but has been adapted for GBM and other instruments. For each burst, selections of the detectors and data types to be used for the catalog analysis were performed by the user based on uniform criteria as described in the main text. The remainder of this appendix describes the procedure and some caveats.

\begin{figure}
\includegraphics[scale=0.75]{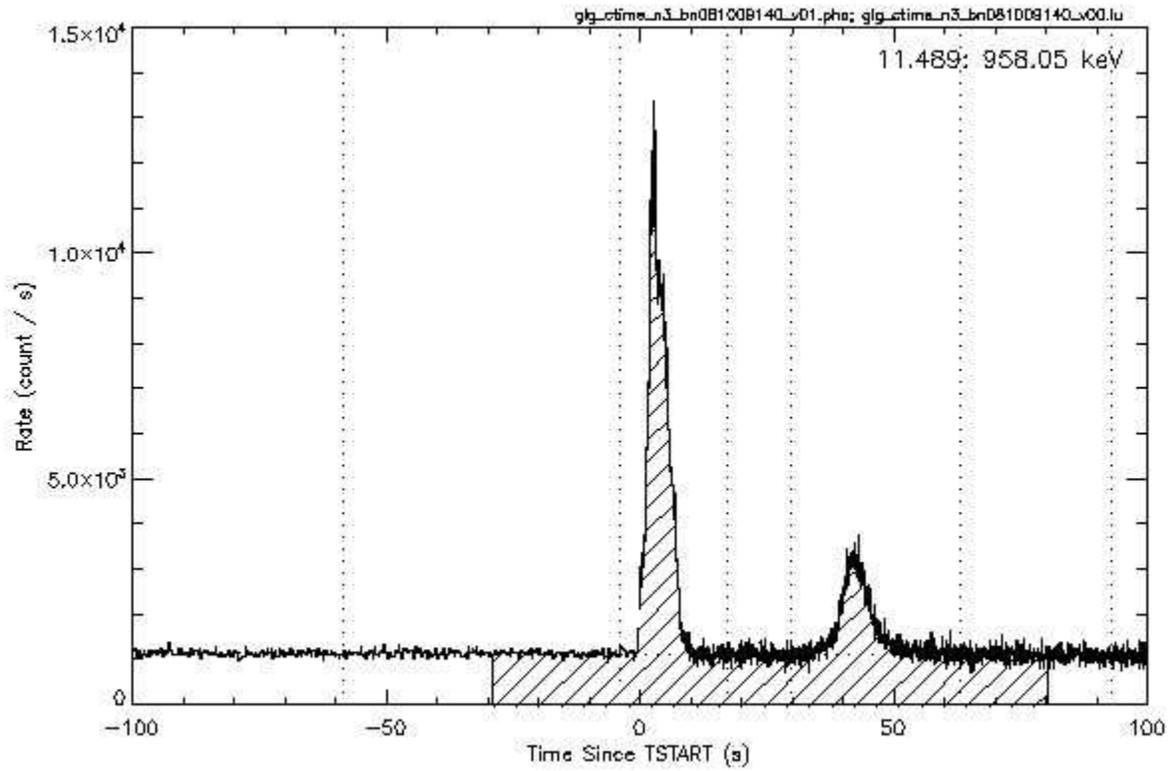}
\caption{\label{duration_lc} CTIME lightcurve of GRB 081009A (bn081009140) in NaI detector~3. Vertical dotted lines indicate the regions selected for fitting the background. Cross-hatching defines the source region selected for the duration analysis. Note that the temporal resolution of CTIME data changes from 0.256~s to 0.064~s at the trigger time.}
\end{figure}

After the data have been selected for a given GRB and read into rmfit, a background model, polynomial in time, is computed separately for each detector, based on user-selected time intervals. The intervals normally include sections earlier and later than the evident burst emission, such that the background model can be interpolated through the entire time of burst activity. In some cases, the background selection includes quiescent portions between pulses of the GRB, if such regions are clearly identifiable. As a final preparation step, the interval for the duration analysis is chosen to cover the entire burst emisson in time, as well as overlapping with the background selections. Figure~\ref{duration_lc} is an example CTIME light curve for GRB 081009A, showing the background and source selections used for the duration analysis.

The next step involves joint spectral fits using the selected detectors for each time bin in the 
selection, including the background regions before and after the burst.
The user is prompted to input a default set of photon model parameters that are 
used when one or both of the spectral shape parameters is undetermined by the 
data in an individual fit, which often occurs for those spectra in the selection region 
that consist of background intervals with the background model subtracted. In 
those cases where the model parameters tend toward unbelievable values, we fix 
that parameter to the default value and redo the fit.  
A poor estimate of the residual spectrum may result in very poor 
spectral fits, dominated by the default model parameters, so the selection of these values may be crucial, especially for weak bursts. In many cases, the duration estimate is more robust when the 
default parameters are set to values that are representative of the background itself (typical values are $E_{\rm peak} = 70$~keV, $\alpha = -1.0$). Although this improves the duration analysis, it may introduce additional systematic errors in the computation of the fluences and peak fluxes.

The choice of photon model to fit is 
dictated by the sparse data statistics: the GBM TTE default time binning and CTIME post-trigger 
accumulation interval is 64 ms, resulting in average source counts that are an order of 
magnitude less than usually required for high-quality spectroscopy ($> 45 \sigma$). The model 
chosen for the catalog analysis is an exponentially cut-off power law, parameterized such that 
the characteristic energy ($E_{\rm peak}$) is identical with the peak in $\nu F_{\nu}$ (the 
so called ``COMP'' model in RMFIT). This model lacks a non-thermal high-energy power-law, 
which is ideal, since it is precisely that parameter that would be least constrained by the 
sparse data at high energies. At the same time, it is desirable to constrain the three 
model parameters that describe the COMP photon model: amplitude, power-law index and 
$E_{\rm peak}$. Hence it is preferable to use datatypes that have few energy channels, so that there are better 
statistics in the channels at hand. CTIME, with 8 energy channels, is usually the best choice, but the pre-trigger time resolution of 256~ms is a limitation for most short bursts as well as a few longer ones. In those cases TTE data may be preferred. 
Native TTE data have the full 128 channels available; however, for the duration analysis these are usually summed to match the energy channels of the corresponding CTIME data.\footnote{For the duration analysis separate software is used to produce a new datatype with 8 energy channels, called CTTE, that is then read into rmfit. This is necessary because by design rmfit always uses the raw energy binning of the input data for spectral fitting.} 
The data are fitted to the available CTIME energy channels, which cover the approximate energy range 10--1000~keV. 

The goal in this analysis is to produce uncertainties in the flux determinations that are no worse than the 
statistical uncertainties due to the counts, while retaining the advantages of correcting for the detector response, 
which is only possible by doing a spectral fit. Just as important, we do not have to reproduce a detailed temporal spectral analysis 
for each burst, which would require summing the data over time until a significant sample has been accumulated 
and would also require better spectral resolution and more complicated spectral models. Instead, we only require that the 
spectral fit in each time bin be reasonably accurate over the energy bins used for the duration calculation (50--300~keV). 
Thus, it is unlikely that the presence of hard emission (such as a high-energy power law, as opposed to our choice of 
an exponential roll-off at high energies) in the data would make a significant contribution to the flux integral. The 
duration energy range is covered by the maximum in the detector response, so the spectral fit is best determined for
those energies. 
The flux uncertainty 
is calculated from the covariance matrix of the fit, so sparse data, such as the background-subtracted background time bins, result 
in uncertainties that are dominated by the best-determined value: the amplitude, which in this case, is driven quite accurately 
to zero. Finally the fitted spectra are further constrained by the required consistency in the joint fit using data from several detectors. 
The spectral parameters and goodness-of-fit for the spectral fit to each time bin may be found in the catalog data files. For most bursts,
the mean goodness-of-fit per degree of freedom is quite consistent with each fit over the entire set of time intervals fitted, indicative of 
normal statistics. For some bright bursts, the model may not be adequate, resulting in higher values of the fit statistic.
However, the excess residuals are typically outside the 50-300 keV energy range, so the effect on the duration calculations is minor. 

\begin{figure}
\begin{center}
\epsscale{1.1}
\plotone{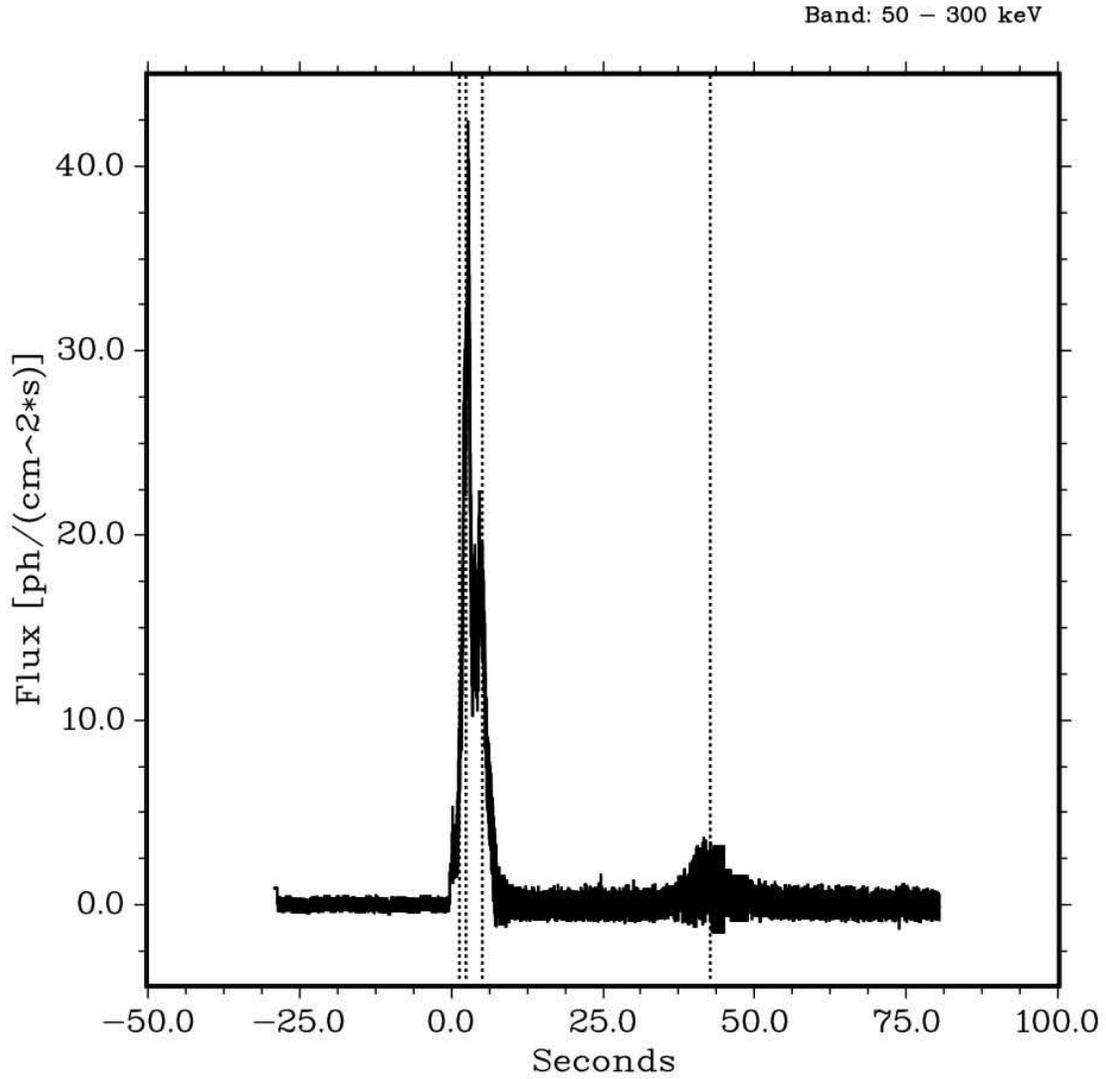}
\caption{\label{flux_lc} Photon flux lightcurve for GRB 081009A (bn081009140) produced by the duration analysis. Data from NaI detectors 3, 4, 7 \& 8 were used. Temporal resolution is the same as in the raw CTIME data. Vertical dotted lines are as described in the caption of Figure~\ref{really_good_duration_plot}.}
\end{center}
\end{figure}

\begin{figure}
\begin{center}
\epsscale{1.1}
\plotone{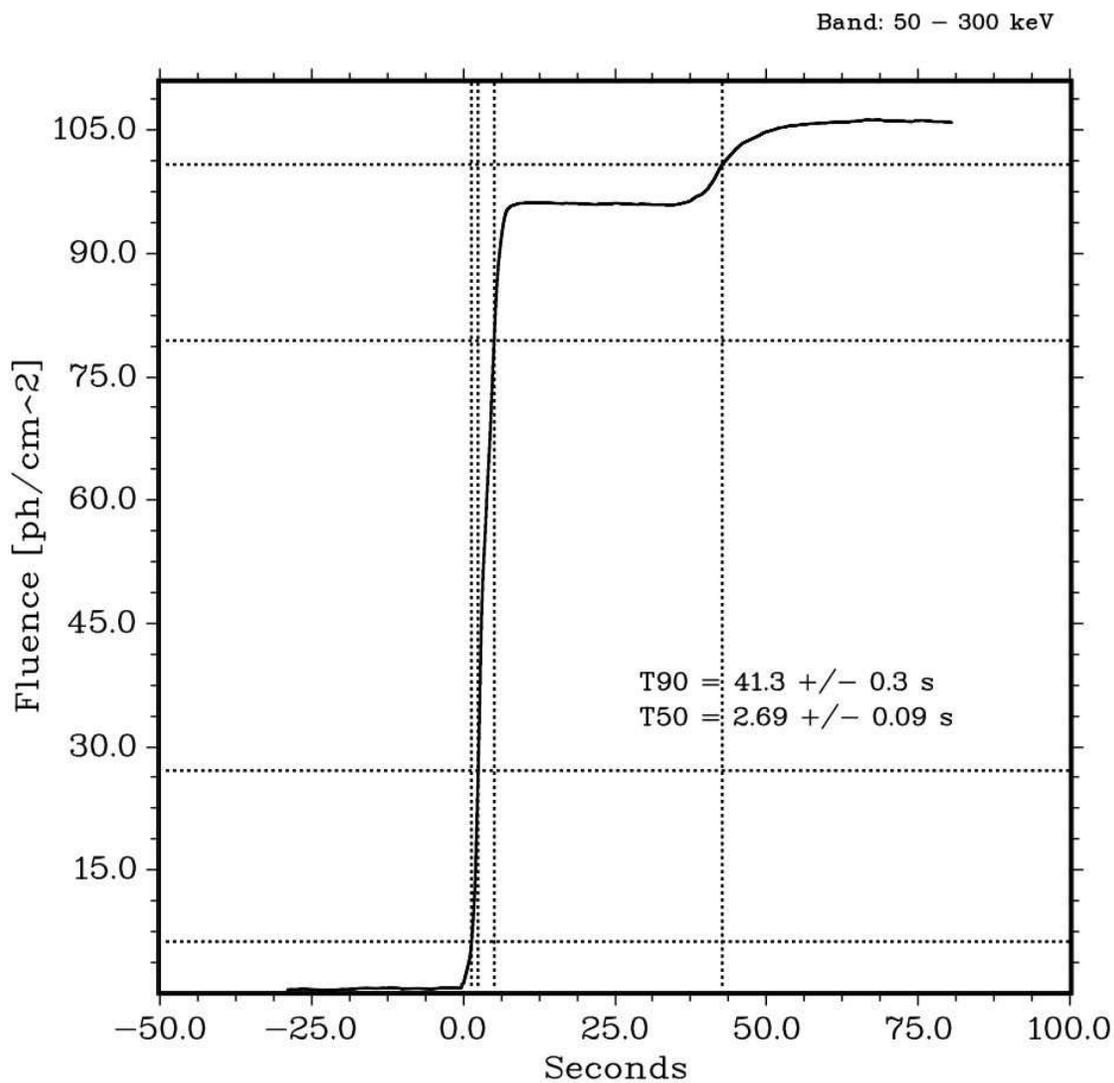}
\caption{\label{really_good_duration_plot} The duration plot for GRB 081009A (bn081009140) is an example of the analysis for a bright GRB. Data from NaI detectors 3, 4, 7 \& 8 were used. Horizontal dotted lines are drawn at 5\%, 25\%, 75\% and 95\% of the total fluence. Vertical dotted lines are drawn at the times corresponding to those same fluences, thereby defining the $T_{50}$ and $T_{90}$ intervals.} 
\end{center}
\end{figure}

\begin{figure}
\begin{center}
\epsscale{1.1}
\plotone{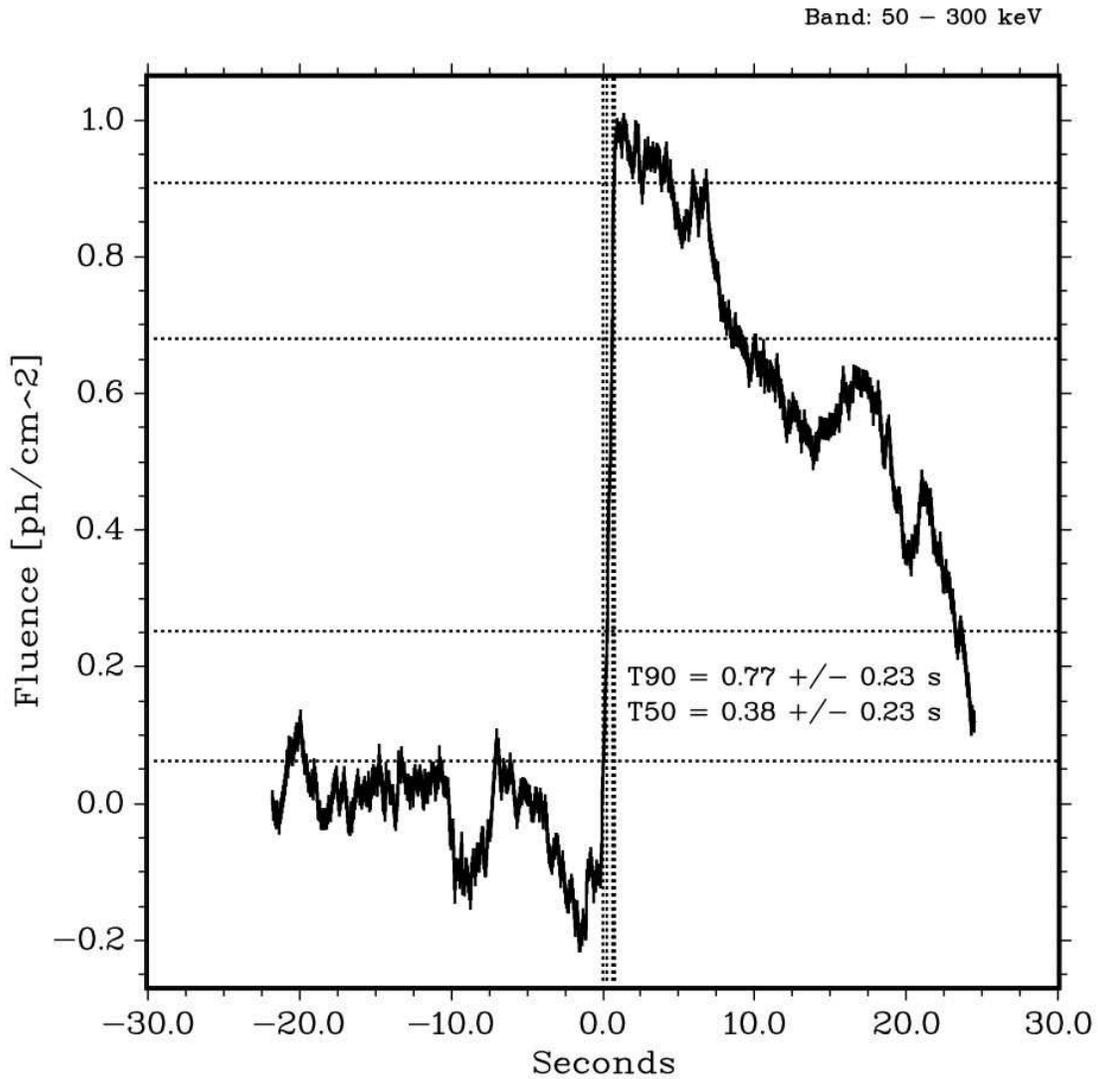}
 \caption{\label{really_awful_plot} The duration plot for GRB 090531B (bn090531775) is an example of the analysis for a weak GRB. Data from NaI detectors 6, 7 \& 9 were used. Dotted lines are as described in the caption for Figure~\ref{really_good_duration_plot}.}
\end{center}
\end{figure}

After the background, source and model parameter selections have been made, every time bin in 
the selection is background subracted, fitted using the model and the resulting fitted spectrum 
is integrated over the appropriate energies to obtain a flux history. For comparison with the BATSE duration distribution, the energy range of 50--300 
keV was chosen for this catalog. Errors for each integrated photon flux are derived using the covariance matrix for the fit, 
taking into account the uncertainties of each fitted model parameter. The resulting photon flux 
history (see, e.\,g., Figure~\ref{flux_lc}) is summed over time, to produce a cumulative
fluence plot, as seen in Figure~\ref{really_good_duration_plot}. In this plot, the 
background-subtracted background 
intervals should, on the average, contribute zero to the total fluence, as seen in the left-hand 
portion of the figure, at times before the trigger time, and at the right-hand portion, well 
after the the burst has concluded. In reality, depending on how well the seeded model fit 
parameters match with the fitted residuals, the flux histories in these two regions can exercise 
a random walk away from constant zero residual flux, as seen in Figure~\ref{really_awful_plot}. 
Similar trends are present in the data in Figure~\ref{really_good_duration_plot} but they are small relative to the burst and hence not noticeable.
In most cases, the random walk over the background accumulation does not exceed 1~ph-cm$^{-2}$, setting a 
practical limit of $> 2$~ph-cm$^{-2}$ in total fluence for the duration analysis to be reliable. Treating this as a hard threshold would, however, bias the catalog against short bursts, which often have lower fluences, so results for weaker bursts are included herein but should be treated with caution.

To 
calculate the duration, the two `plateau' regions must be identified by hand (since every such 
plot is unique, this step can not be automated); the average flux in each serving as the fiducial 
values against which the partial fluences at 5, 25, 75 and 95\% of total are determined. At each 
fraction of the total fluence, its intersection with the integrated flux history is projected 
vertically onto the time axis, giving four time values, $t_5, t_{25}, t_{75}$ and $t_{95}$. 
The duration measure is defined as: $T_{90} = t_{95} - t_{5}$, or the interval between the times 
where the burst has reached 5\% and 95\% of its total emission in the 50--300 keV band. As 
shown by \citet{Koshut}, the robustness of the $T_{90}$ estimate for the duration relies upon the 
integrated flux history being single-valued at the two fractional fluence values. This, in turn, 
implies that the fluence levels at $t_5$ and $t_{95}$ should be somewhat larger than the 
variance of the corresponding nearest plateau region; otherwise, there is confusion as to 
which time to be used to identify each of these. $T_{50} = t_{75} - t_{25}$ is based upon 
flux levels that are presumably further away from the random-walk levels, and thus is considered 
to be more robust than $T_{90}$.


Following \citet{Koshut}, the variance of the two residual plateau intervals is used as a basis 
for the error estimates for $T_{90}$ and $T_{50}$. For that reason, the 
plateau regions are chosen to
contain enough samples that their variance is representative of the
residual fluctuations in the background-subtracted background time bins. Our estimation of the 
background evolution in time is based upon a polynomial fit over user-selected regions before and 
after the GRB emission start and end times; for the best results, the plateau selection should 
overlap with the regions selected for the fitting of the background. Ideally, the residuals from the background 
fit should then be zero in the region where the plateau selections overlap the background selections, 
but in practice the background-subtracted time bins in the background regions have fluctuations that can be as large as 1~ph-cm$^{-2}$. These small, higher-order fluctuations drive the uncertainty calculation for the 
flux levels, since the variance measures our inability to precisely determine the zero and 100\% 
levels.
Once the variances in the flux histories are known, they are converted into 
uncertainties by first scaling by the desired flux level and taking the square root: 
\begin{equation}
error_{\rm flux_{nn}} = \sqrt{(1- nn)^2 Var_0 + nn^2 Var_{100}},
\end{equation}
where $nn \in \{0.05, 0.25, 0.75, 0.95\}$ indicates the various flux levels and $Var_0$ and 
$Var_{100}$ are the variances from the zero and 100\% fluence level plateaus. For each of 
the four nominal fluence levels $nn$, $error_{\rm flux_{nn}}$ is added and subtracted, resulting 
in a projected uncertainty in time for each. The final uncertainty in $T_{90}$ is the root-mean-square of
the corresponding uncertainties in $t_5$ and $t_{95}$ and similarly with the projected 
uncertainties in $t_{25}$ and $t_{75}$ for $T_{50}$. The start time, relative 
to the burst trigger time, is also recorded for each of the time intervals that form $T_{90}$ and $T_{50}$.

The flux history used in the calculation of the burst duration can be used to derive several 
other important quantities. 
The total fluence is calculated by differencing the zero and 100\% integrated flux levels, as determined 
by the plateau selections. In each successive time bin the flux model from the fit is 
weighted by energy in erg, integrated over two energy bands, 50--300~keV and 10--1000~keV, and then 
added to the running total to produce the cumulative fluence. The variances of the two plateau regions are 
added together in quadrature to determine the uncertainty in the fluence. As with the $T_{90}$ 
calculation, it is the variance in the background regions (where the running sum should be zero) 
that determines the uncertainty of the zero level fluence (and similarly for the total fluence).
Given the limited range of integration for the 50-300 keV fluences, the fact that we chose to use the COMP photon
model makes little difference to the results; however, this may have a more significant effect on the 10-1000 keV fluences. 
The affected bursts are primarily the ones with significant emission in the BGOs, which are highlighted in Table 2. In any case, the fluences derived in this catalog are intended mostly as a ranking tool. Readers interested in more robust fluence estimates 
should consult the GBM spectroscopy catalog \citep{spec_cat}. 

The deconvolved photon flux history is calculated by integrating the best fit model for each time bin 
over the two energy bands described above. The peak flux is then the maximum value of the  
flux history between the lower and upper plateaus for the two energy 
bands, as well as for three different time intervals: 64, 256 and 1024 ms. 
As the native or default accumulation for the CTIME data post-trigger is 64 ms, there is only 
one possibility for binning, as long as care is taken to ensure that the peak flux interval 
occurred after $T = 0$, as is usually the case for long GRBs. For short GRBs, TTE 
is preferred, since much of the emission can occur pre-trigger, and TTE can 
be binned in 64~ms accumulations over its entirety. In order to calculate the 256 and 1024~ms 
peak fluxes, the available data are binned within a sliding window. CTIME pre-trigger 
accumulations are 256~ms by default, so only the post-trigger data need to be binned. In this 
case (CTIME), the peak flux is the maximum flux found either pre-trigger or in one of the sliding 
binning windows post-trigger. The time and value of the peak flux and its uncertainty are recorded, 
again calculated from the model fit, along with the uncertainties of the model parameters and the covariance matrix between the parameters, computed in the usual manner.

\clearpage

 \clearpage
 
 \clearpage
  
 \clearpage

\clearpage
 
\clearpage

 \clearpage
 
\clearpage

\clearpage

\clearpage

\clearpage

\clearpage




\begin{deluxetable}{cccccccccc}

\rotate

\tabletypesize{\small}

\tablewidth{614pt}

\tablecaption{\label{trigger:criteria:history} Trigger Criteria HIstory}


\tablehead{\colhead{Algorithm} & \colhead{Timescale} & \colhead{Offset} & \colhead{Channels} & \colhead{Energy} & \multicolumn{5}{c}{Threshold ($0.1 \sigma$)\tablenotemark{a}}\\ 
\cline{6-10} \\
\colhead{Number} & \colhead{(ms)} & \colhead{(ms)} & \colhead{} & \colhead{(keV)} & \colhead{2008-07-11} & \colhead{2008-07-14} & \colhead{2008-08-01} & \colhead{2009-05-08} & \colhead{2009-07-02} } 

\startdata
1 & 16 & 0 & 3--4 & 50--300 & 75 & : & : & : & : \\
2 & 32 & 0 & 3--4 & 50--300 & 75 & : & : & : & : \\
3 & 32 & 16 & 3--4 & 50--300 & 75 & : & : & : & : \\
4 & 64 & 0 & 3--4 & 50--300 & 45 & : & 50 & : & : \\
5 & 64 & 32 & 3--4 & 50--300 & 45 & : & 50 & : & : \\
6 & 128 & 0 & 3--4 & 50--300 & 45 & : & 48 & 50 & : \\
7 & 128 & 64 & 3--4 & 50--300 & 45 & : & 48 & 50 & : \\
8 & 256 & 0 & 3--4 & 50--300 & 45 & : & : & : & : \\
9 & 256 & 128 & 3--4 & 50--300 & 45 & : & : & : & : \\
10 & 512 & 0 & 3--4 & 50--300 & 45 & : & : & : & : \\
11 & 512 & 256 & 3--4 & 50--300 & 45 & : & : & : & : \\
12 & 1024 & 0 & 3--4 & 50--300 & 45 & : & : & : & : \\
13 & 1024 & 512 & 3--4 & 50--300 & 45 & : & : & : & : \\
14 & 2048 & 0 & 3--4 & 50--300 & 45 & : & : & : & : \\
15 & 2048 & 1024 & 3--4 & 50--300 & 45 & : & : & : & : \\
16 & 4096 & 0 & 3--4 & 50--300 & 45 & : & : & : & : \\
17 & 4096 & 2048 & 3--4 & 50--300 & 45 & : & : & : & : \\
18 & 8192 & 0 & 3--4 & 50--300 & C & 50 & : & : & D \\
19 & 8192 & 4096 & 3--4 & 50--300 & C & 50 & : & : & D \\
20 & 16384 & 0 & 3--4 & 50--300 & C & 50 & D & : & : \\
21 & 16384 & 8192 & 3--4 & 50--300 & C & 50 & D & : & : \\
22 & 16 & 0 & 2--2 & 25--50 & D & 80 & : & : & : \\
23 & 32 & 0 & 2--2 & 25--50 & D & 80 & : & : & : \\
24 & 32 & 16 & 2--2 & 25--50 & D & 80 & : & : & : \\
25 & 64 & 0 & 2--2 & 25--50 & D & 55 & : & : & : \\
26 & 64 & 32 & 2--2 & 25--50 & D & 55 & : & : & : \\
27 & 128 & 0 & 2--2 & 25--50 & D & 55 & : & : & D \\
28 & 128 & 64 & 2--2 & 25--50 & D & 55 & : & : & D \\
29 & 256 & 0 & 2--2 & 25--50 & D & 55 & : & : & D \\
30 & 256 & 128 & 2--2 & 25--50 & D & 55 & : & : & D \\
31 & 512 & 0 & 2--2 & 25--50 & D & 55 & : & : & D \\
32 & 512 & 256 & 2--2 & 25--50 & D & 55 & : & : & D \\
33 & 1024 & 0 & 2--2 & 25--50 & D & 55 & : & : & D \\
34 & 1024 & 512 & 2--2 & 25--50 & D & 55 & : & : & D \\
35 & 2048 & 0 & 2--2 & 25--50 & D & 55 & : & : & D \\
36 & 2048 & 1024 & 2--2 & 25--50 & D & 55 & : & : & D \\
37 & 4096 & 0 & 2--2 & 25--50 & D & 65 & : & : & D \\
38 & 4096 & 2048 & 2--2 & 25--50 & D & 65 & : & : & D \\
39 & 8192 & 0 & 2--2 & 25--50 & D & 65 & : & : & D \\
40 & 8192 & 4096 & 2--2 & 25--50 & D & 65 & : & : & D \\
41 & 16384 & 0 & 2--2 & 25--50 & D & 65 & D & : & : \\
42 & 16384 & 8192 & 2--2 & 25--50 & D & 65 & D & : & : \\
43 & 16 & 0 & 5--7 & $> 300$ & D & 80 & : & : & : \\
44 & 32 & 0 & 5--7 & $> 300$ & D & 80 & : & : & D \\
45 & 32 & 16 & 5--7 & $> 300$ & D & 80 & : & : & D \\
46 & 64 & 0 & 5--7 & $> 300$ & D & 55 & : & 60 & D \\
47 & 64 & 32 & 5--7 & $> 300$ & D & 55 & : & 60 & D \\
48 & 128 & 0 & 5--7 & $> 300$ & D & 55 & : & : & D \\
49 & 128 & 64 & 5--7 & $> 300$ & D & 55 & : & : & D \\
50 & 16 & 0 & 4--7 & $> 100$ & D & 80 & : & : & : \\
51 & 32 & 0 & 4--7 & $> 100$ & D & 80 & : & : & D \\
52 & 32 & 16 & 4--7 & $> 100$ & D & 80 & : & : & D \\
53 & 64 & 0 & 4--7 & $> 100$ & D & 55 & : & : & D \\
54 & 64 & 32 & 4--7 & $> 100$ & D & 55 & : & : & D \\
55 & 128 & 0 & 4--7 & $> 100$ & D & 55 & : & : & D \\
56 & 128 & 64 & 4--7 & $> 100$ & D & 55 & : & : & D \\
57 & 256 & 0 & 4--7 & $> 100$ & D & 55 & : & : & D \\
58 & 256 & 128 & 4--7 & $> 100$ & D & 55 & : & : & D \\
59 & 512 & 0 & 4--7 & $> 100$ & D & 55 & : & : & D \\
60 & 512 & 256 & 4--7 & $> 100$ & D & 55 & : & : & D \\
61 & 1024 & 0 & 4--7 & $> 100$ & D & 55 & : & : & D \\
62 & 1024 & 512 & 4--7 & $> 100$ & D & 55 & : & : & D \\
63 & 2048 & 0 & 4--7 & $> 100$ & D & 55 & : & : & D \\
64 & 2048 & 1024 & 4--7 & $> 100$ & D & 55 & : & : & D \\
65 & 4096 & 0 & 4--7 & $> 100$ & D & 65 & : & : & D \\
66 & 4096 & 2048 & 4--7 & $> 100$ & D & 65 & : & : & D \\
\enddata

\tablenotetext{a}{ Symbol ':' indicates no change from previous setting; 'C' indicates that the algorithm is in compute mode (see text); 'D' indicates that the algorithm is disabled.}



\end{deluxetable}
 



 commands
\tablenotetext{a}{Bursts with Trigger ID and GRB Name in italics have significant emission in at least one BGO detector (see text).}
\tablenotetext{b}{Other instrument detections: K: Konus-WInd, K-RF: Konus-RF, S: Swift, IA: INTEGRAL SPI-ACS, IB: INTEGRAL Burst Alert System, W: Suzaku-WAM, R: RHESSI, M: MAXI, SA: SuperAGILE, AM: AGILE-MCAL, A: AGILE, L: Fermi LAT}
\tablenotetext{c}{GRB091024A triggered GBM twice.}



\end{deluxetable}




 commands
\tablenotetext{a}{Data problems precluded duration analysis.}
\tablenotetext{b}{Used TTE binned at 32 ms.}
\tablenotetext{c}{Partial earth occultation is likely; durations are lower limits.}
\tablenotetext{d}{Possible precursor at $\sim T_0-120$ s.}
\tablenotetext{e}{Data cut off while burst in progress; durations are lower limits.}
\tablenotetext{f}{SAA entry at $T_0+83$~s; durations are lower limits.}
\tablenotetext{g}{GRB091024 triggered GBM twice.}
\tablenotetext{h}{Too weak to measure durations; visual duration is $\sim 0.025$~s.}
\tablenotetext{i}{Possible contamination due to emergence of Crab \& A0535+26 from Earth occultation.}
\tablenotetext{j}{Used TTE binned at 16 ms.}



\end{deluxetable}




 commands



\end{deluxetable}

\end{document}